\PassOptionsToPackage{english}{babel}

\documentclass[aps,pra,reprint,superscriptaddress,letterpaper,twocolumn,longbibliography,floatfix]{revtex4-1}
\usepackage{graphicx}
\usepackage{amssymb}     
\usepackage[english,ngerman]{babel}

\usepackage{blindtext}
\usepackage{amsmath} 
 \usepackage[T1]{fontenc}

\usepackage{mathtools}
\usepackage{cleveref}

\newmuskip\pFqmuskip

\newcommand*\pFq[6][8]{%
  \begingroup 
  \pFqmuskip=#1mu\relax
  \mathcode`\,=\string"8000
  \begingroup\lccode`\~=`\,
  \lowercase{\endgroup\let~}\pFqcomma
  {}_{#2}F_{#3}{\left[\genfrac..{0pt}{}{#4}{#5};#6\right]}%
  \endgroup
}
\newcommand{\pFqcomma}{\mskip\pFqmuskip}

\usepackage{setspace}
\usepackage[usenames, dvipsnames, x11names]{xcolor}

\def    \bse{\begin{subequations}}
\def    \ese{\end{subequations}}

\newcommand{\pd}[2]{\frac{\partial #1}{\partial #2}} 
\newcommand{\pdd}[2]{\frac{\partial^2 #1}{\partial #2^2}} 
\newcommand{\ket}[1]{\left| #1 \right>} 
\newcommand{\bra}[1]{\left< #1 \right|} 
\newcommand{\braket}[2]{\left< #1 \vphantom{#2} \right|
 \left. #2 \vphantom{#1} \right>} 
\let\baraccent=\= 
\renewcommand{\=}[1]{\stackrel{#1}{=}} 

\providecommand{\fr}{\frac}

\begin{document}

\selectlanguage{english}

\title{Driven-dissipative quantum Kerr resonators: new exact solutions, photon blockade and quantum bistability}

\author{David Roberts}
\affiliation{Department of Physics, University of Chicago, Chicago, IL 60637, USA}
\affiliation{Pritzker School  for  Molecular  Engineering,  University  of  Chicago, 5640  South  Ellis  Avenue,  Chicago,  Illinois  60637,  U.S.A.}

\author{Aashish A. Clerk}
\affiliation{Pritzker School  for  Molecular  Engineering,  University  of  Chicago, 5640  South  Ellis  Avenue,  Chicago,  Illinois  60637,  U.S.A.}

\date{\today}

\begin{abstract}  
We present a new approach for deriving exact, closed-form solutions for the steady state of a wide class of driven-dissipative nonlinear resonator that is distinct from more common complex-$P$ function methods.  Our method generalizes the coherent quantum absorber approach of Stannigel et al. \cite{stannigel_driven-dissipative_2012} to include nonlinear driving and dissipation, and relies crucially on exploiting the Segal-Bargmann representation of Fock space.  Our solutions and method reveal a wealth of previously unexplored observable phenomena in these systems, including new generalized photon-blockade and anti-blockade effects, and an infinite number of new parameter choices that yield quantum bistability.  \end{abstract}

\maketitle

\section{Introduction}

Exact solutions of interacting, driven-dissipative quantum problems are rare, and thus occupy a special place in the study of open quantum systems.  A canonical example is the solution of the driven-dissipative Kerr resonator.  Here, a bosonic mode with a Kerr nonlinearity (i.e.~a Hubbard $U$ interaction) is subject to a coherent linear drive and Markovian single photon loss.  As shown by Drummond and Walls \cite{drummond_quantum_1980}, one can exactly solve for the steady state of this system using a complex-$P$ phase space representation. Later work showed that models including two-photon driving and loss are also solvable using this technique \cite{drummond_quantum_1980,bartolo_exact_2016, elliott_applications_2016}.  
These driven nonlinear cavity systems have renewed relevance, as they can be directly implemented in superconducting circuit QED setups (see, e.g.,  \cite{Kirchmair2013,leghtas_confining_2015,touzard_coherent_2018,Leghtas2019,Grimm2019}).  Their ability to exhibit multiple steady states has utility in quantum information processing \cite{mirrahimi_dynamically_2014,Goto2016,puri_engineering_2017}.  

While the existence of exact solutions here are remarkable, they are somewhat physically opaque and unwieldy (e.g.~they are typically expressed as infinite sums of special functions).  Their derivation is also somewhat intricate, requiring a non-trivial integration to relate the solution of an effective classical problem to the underlying quantum system.  More direct methods for obtaining and possibly extending these solutions are thus highly desirable.  For the simplest version of the Kerr-cavity problem (single-photon drive and loss only), Stannigel et al.~\cite{stannigel_driven-dissipative_2012} were able to reproduce the exact solution of Ref.~\cite{drummond_quantum_1980} using a simple, purely algebraic approach.  While extremely elegant, it was unclear whether this approach could be extended to more complex problems.
 \begin{figure}[!t]
     \centering
    \includegraphics[width=0.95\columnwidth]{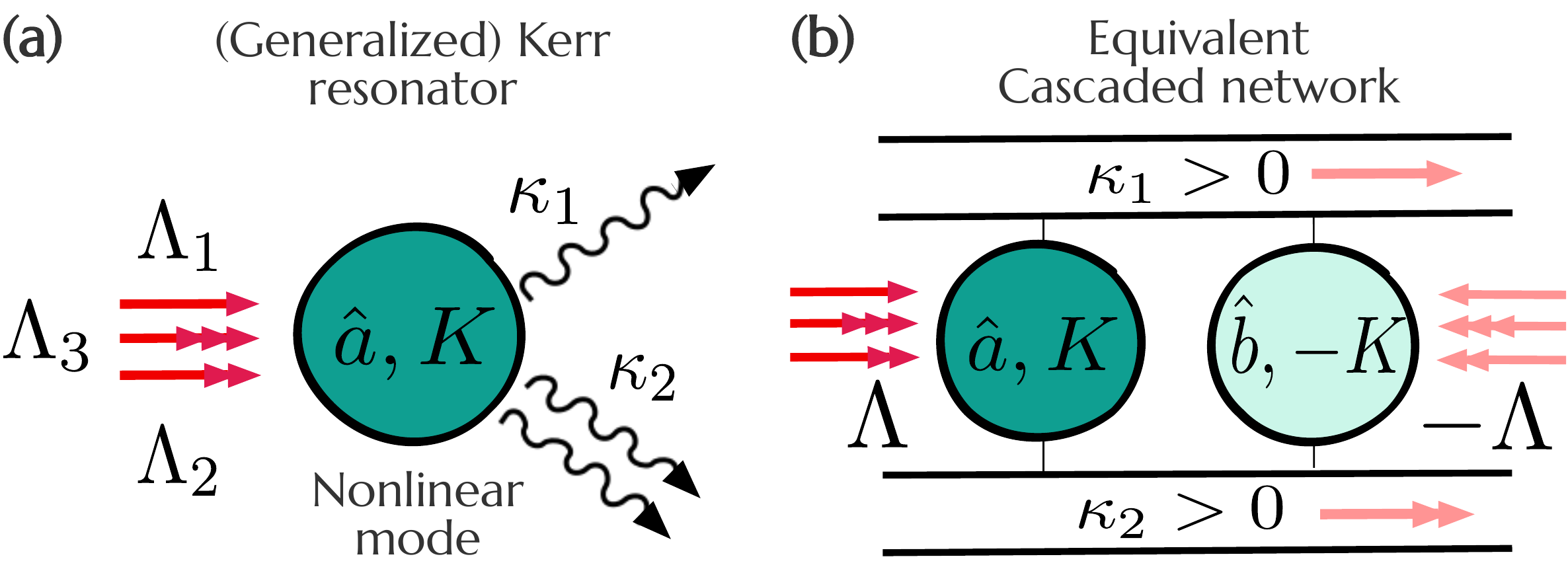}
     \caption{(a) Generalized driven Kerr cavity problem, where a single interacting bosonic mode is subject to linear and nonlinear coherent drives $\Lambda_j$, as well as independent one and two photon loss (rates $\kappa_1$, $\kappa_2$).  (b)  The coherent quantum absorber (CQA) method represents each dissipative bath as a chiral waveguide, and couples a second auxiliary $b$ cavity downstream.  
     By picking its Hamiltonian judiciously, the entire composite system can relax to a pure state, providing an efficient means for finding the steady state of cavity-$a$.}
\label{fig:virtual_cavity_intro}
 \end{figure} 

In this paper, we show that such an extension is indeed possible:  the 
``coherent quantum absorber'' (CQA) method
of Ref.~\cite{stannigel_driven-dissipative_2012}
can be extended to a wide class of systems which include nonlinear coherent driving as well as multiple dissipators (see Fig.~\ref{fig:virtual_cavity_intro}).  Our extension employs a new ingredient:  the Segal-Bargmann representation of a single-mode pure state wavefunction \cite{bargmann_hilbert_1961,Segal1962,Segal1963,bargmann_hilbert_1967}. This enables non-trivial transformations that are crucial for finding exact solutions.  
Our approach yields several new insights.  We find and describe new parameter regimes where the steady-state exhibits a surprising generalized photon-blockade phenomenon. In particular, we show how the use of a nonlinear driving term allows for photon blockade even for nonlinearities much weaker than dissipation rates; unlike so-called ``unconventional photon blockade" \cite{liew_single_2010,bamba_origin_2011,lemonde_antibunching_2014}, the effect we describe results in non-Gaussian states and a complete suppression of higher-$n$ photon probabilities.

We also find an infinite number of points in parameter space where our generalized driven Kerr system exhibits quantum bistability (i.e.~a two-dimensional decoherence-free subsystem), despite the lack of photon-number parity conservation.  The required parameters can be achieved asymptotically in the limit of weak single-photon loss.  Our solution also 
provides a simple intuitive picture when there is a unique steady state (see 
Fig.~\ref{fig:beam_splitter}):  the steady state is formed by mixing a pure state with vacuum at a 50-50 beamsplitter, and then discarding one of the outputs.
At a technical level, our work also provides new, simple closed-form expressions for the steady-state Wigner function and normally-ordered moments.

The remainder of this paper is organized as follows.  
In Sec.~\ref{sec:system} we introduce the basic system.  In Sec.~\ref{sec:the absorber method}, we review the
CQA method, and in Sec.~\ref{sec:soln_of_general_kerr} present our extension to nonlinear driving and multiple dissipators.   Sec.~\ref{sec:Insights} summarizes the new physical phenomena uncovered by our exact solution, while Sec.~\ref{sec:nearby} and Sec.~\ref{sec:bistability} discuss regimes of classical and quantum bistability.  We conclude in Sec.~\ref{sec:conclusions}.

\section{System}\label{sec:system}

We consider a driven Kerr resonator whose coherent dynamics is described by the Hamiltonian
\begin{align}
    \hat{H}_a &=
        \fr{K}{2}\hat{a}^\dag \hat{a}^\dag \hat{a}\hat{a} -\Delta \hat{a}^\dag \hat{a}
\nonumber        \\
        &+ \bigg[ 
            \bigg(\Lambda_1\hat{a}^\dag + \fr{\Lambda_2}{2}\hat{a}^\dag \hat{a}^\dag+\Lambda_3 \hat{a}^\dag \hat{a}^\dag \hat{a} \bigg) + h.c. \bigg]  . \label{eq:ciuti_ham}
\end{align}
We work in a rotating frame, and have assumed that all drives have an equal detuning $\Delta$ from the cavity resonance frequency (which allows us to have a time-independent rotating-frame Hamiltonian).  Here $K$ is the Kerr nonlinearity, and $\Lambda_1,\Lambda_2$ are the complex amplitudes of standard coherent one and two photon driving terms.  $\Lambda_3$ represents an unusual kind of nonlinear single-photon driving term; as we will see, it enables a striking new kind of photon blockade effect that does not require strong nonlinearity.  We show in Appendix \ref{app:cQED} how this $\Lambda_3$ drive can be implemented using the superconducting circuit architecture of Ref. \cite{sivak_kerr_2019}.

The full dissipative dynamics includes one and two photon loss processes, and is described by the Lindblad master equation
\begin{align}
    \frac{d}{dt} \hat{\rho} 
        &=-i[\hat{H}_a,\hat{\rho}] + \kappa_1 \mathcal{D} [\hat{a}] \hat{\rho} +
        \kappa_2 \mathcal{D} [\hat{a}^2 ] \hat{\rho} \equiv 
        \mathcal{L}_0  \hat{\rho},
    \label{eq:one_mode}
\end{align}
where $\mathcal{D}[\hat{X}] \hat \rho \equiv \hat{X} \hat{\rho} \hat{X}^\dagger - (1/2) \left \{ \hat{X}^\dagger \hat{X}, \hat{\rho} \right \}$ is the usual Lindblad dissipative superoperator, and $\kappa_1$ ($\kappa_2$) are the one (two) photon decay rates.  Note that the dissipative evolution corresponds to coupling the system to two distinct zero-temperature baths.

We will focus exclusively on finding the steady states of this kind of system, 
i.e.~density matrices $\hat{\rho}_{\rm ss}$ satisfying
\begin{equation}
    \mathcal{L}_0 \hat{\rho}_{\rm ss} = 0.
\end{equation}
We briefly summarize prior work on this model. For $\Lambda_3\equiv 0$, exact solutions for $\hat{\rho}_{\rm ss}$ have been found using the complex $P$-function approach \cite{drummond_quantum_1980,drummond_non-equilibrium_1981, elliott_applications_2016,bartolo_exact_2016}.  The solutions express matrix elements of $\hat{\rho}_{\rm ss}$ in the Fock basis as sums of special functions.  In the semiclassical limit, solutions for the steady state can be found using an alternate approach developed
by Dykman and co-workers \cite{marthaler_switching_2006,dykman_periodically_2012}; unlike the complex-$P$ approach, these can also be used to describe dissipation at a non-zero temperature. Systems with higher-order coherent driving terms (like our $\Lambda_3$) have been studied previously, (see, e.g., \cite{guo_phase_2013,svensson_period_2018,Lorch2019}), but were not previously known to be solvable.

While the prior work on driven Kerr resonators is a remarkable achievement, it leaves several mysteries unanswered.  First, in the presence of single photon loss, the unique steady state that one finds {\it always} yields a positive-definite Wigner function.  Given the nonlinearity in the system, it is not {\it a priori} obvious that this should necessarily be the case. Second, in the absence of single photon processes (i.e.~$\Lambda_1 = \kappa_1 = 0$), this system exhibits multiple steady states \cite{krippner_transient_1994, wolinsky_quantum_1988,hach_iii_generation_1994,gilles_generation_1994,mirrahimi_dynamically_2014}.  We are not aware of any discussion of this using the complex-$P$ approach.  For $\Delta = 0$, the system is simple enough that the multiple steady states can be found via elementary means, in terms of superpositions of coherent states \cite{Goto2016,puri_engineering_2017}. Conditions needed for Wigner function negativity were recently discussed in Ref. \cite{braasch_wigner_2019}, though these are not directly applicable to our system. In the sections that follow, we discuss an alternate, physically-transparent method for solving this class of problems that addresses the open issues mentioned above.

\section{Exact solutions using the quantum absorber method}
\label{sec:the absorber method}

Our approach to solving driven-dissipative Kerr problems is to adapt and extend the so-called ``coherent quantum absorber'' (CQA) approach first introduced by Stannigel et 
al.~\cite{stannigel_driven-dissipative_2012} to solve the simplest driven Kerr problem where there are no two photon drive or loss processes.  We quickly recap the philosophy of this approach, and then show how it can be extended to deal with more complex problems involving two and even three photon processes.

\subsection{Recap of the basic approach}

Consider first the case where our system in Eq.~(\ref{eq:one_mode}) has no two photon loss ($\kappa_2=0$).  The starting point of the CQA method  is to represent the one photon loss as arising from a coupling to a chiral (i.e.~unidirectional) waveguide.  Further, one imagines coupling a second auxiliary bosonic mode (annihilation operator $\hat{b}$, system Hamiltonian $\hat{H}_b$) to the waveguide, downstream from the physical $a$ cavity (see Fig.~\ref{fig:virtual_cavity_intro}).  Given the chirality of the waveguide, the dynamics of this auxiliary cavity can have no impact on the physical cavity $a$.  The entire composite system can be described using standard cascaded quantum systems theory \cite{carmichael_quantum_1993,gardiner_driving_1993,gardiner_quantum_2000}.  The dynamics of the reduced density matrix $\hat{\rho}_{ab}$ describing both cavities is described by a Lindblad master equation of the form:
\begin{align}
    \frac{d}{dt} \hat{\rho}_{ab} &= -i[\hat{H}_{ab},\hat{\rho}_{ab}] +\kappa_1 \mathcal{D}[\hat{a}-\hat{b}]\hat{\rho}_{ab},
    \label{eq:two_mode}\\
    \hat{H}_{ab} &= 
        \hat{H}_a + \hat{H}_b 
    -\fr{i\kappa_1}{2}(\hat{a}^\dag \hat{b}-h.c.).
    \label{eq:cascaded_ham}
\end{align}
Note that one can rigorously trace out cavity $b$ from this equation, recovering Eq.~(\ref{eq:one_mode}) for cavity $a$ alone.  

While the introduction of the auxiliary cavity $b$ has no impact on cavity $a$, it provides a useful tool for finding its steady state.  As shown in Ref.~\cite{stannigel_driven-dissipative_2012}, for a general cavity $a$ Lindblad master equation having only single-photon loss (i.e.~Eq.~(\ref{eq:one_mode}) with $\kappa_2 = 0$ and arbitrary $\hat{H}_a$), one can {\it always} construct a Hamiltonian $\hat{H}_b$ for the auxiliary cavity $b$ such that the composite system has a pure steady state.  This steady state state necessarily has vanishing emission to the waveguide--  it is a ``dark'' state.  Letting $\hat{\rho}_{ab,ss}$ denote the steady-state density  matrix of the two-cavity problem, this means:
\begin{align}
    \hat{\rho}_{ab,ss} & = \ket{\psi} \bra{\psi}, \hspace{0.6 cm}
    \left(\hat{a} - \hat{b} \right) \ket{\psi} = 0.
    \label{eq:the absorber methodDarkStateCondition}
\end{align}
Note that the dark state condition implies that $\ket{\psi}$ is essentially a single mode state.  Introducing new composite mode operators
\begin{equation}
    \hat{c}_\pm \equiv \frac{\hat{a}\pm \hat{b}}{\sqrt{2}},
    \label{eq:cmodes}
\end{equation}
one notes that the dark state condition forces the composite mode $\hat{c}_-$ to be in vacuum.  Hence, one just needs to solve for the (pure) state of the composite $\hat{c}_+$ mode.

In physical terms, the CQA approach seeks to construct $\hat{H}_b$ such that the auxiliary cavity $b$ acts as a ``perfect absorber'' for all photons emitted into the waveguide by cavity $a$.  By tracing out cavity $b$, one obtains the desired steady state for the physical cavity-$a$ problem.  One generically obtains an impure state, as the two cavities will be entangled in the state $\ket{\psi}$.

While such a construction is always possible, in practice it would seem to be of no utility, as one can only construct the required $\hat{H}_b$ by first {\it independently} solving for the cavity-$a$ steady $\hat{\rho}_{a,ss}$,  Despite this seeming obstacle, Ref.~\cite{stannigel_driven-dissipative_2012} demonstrated that for a range of problems, one could essentially guess the form of $\hat{H}_b$ without first knowing $\hat{\rho}_{a,ss}$.  This educated guess is extremely simple:  $\hat{H}_b$ is taken to be identical to $\hat{H}_a$ up to an overall minus sign.  
Ref.~\cite{stannigel_driven-dissipative_2012}
applied this to the simplest driven Kerr problem ($\Lambda_2 = \Lambda_3 = \kappa_2 = 0$ in Eq.~(\ref{eq:ciuti_ham})), in which case
\begin{equation}
    \hat{H}_{b} = 
          -\fr{K}{2}\hat{b}^\dag \hat{b}^\dag \hat{b}\hat{b} 
          +\Delta \hat{b}^\dag \hat{b}
        - \left[ 
            \Lambda_1\hat{b}^\dag  + h.c. \right].
\end{equation}
With this choice, Stannigel et al. were able to find a pure-state solution of the cascaded master equation in Eq.~(\ref{eq:two_mode}) by solving a  simple one-term recursion relation.  By then tracing out cavity $b$, they recovered (in a much simpler manner) the classic solution of Drummond et al. \cite{drummond_quantum_1980} for the linear-drive Kerr problem.

\subsection{Extension to nonlinear driving and two-photon loss}

It is natural to ask whether the absorber method approach can be extended to solve problems with nonlinear driving and two-photon loss.  
An immediate issue is the presence of two independent dissipators in the master equation Eq.~(\ref{eq:one_mode}).  We find that the CQA approach is easily modified to deal with this situation.  As shown in Fig.~\ref{fig:virtual_cavity_intro}(b), one can represent the two-photon loss process as a nonlinear coupling to a second chiral waveguide.  

One again needs to add something downstream along this waveguide to absorb the emitted excitations.  While there are many possible options, we find the simplest approach is sufficient:  we assume that there is still a single auxiliary cavity $b$ that now couples to {\it both} these independent chiral waveguides.  The cascaded master equation now takes the form:
\begin{align}
    \frac{d}{dt} \hat{\rho}_{ab} &= 
        -i[\hat{H}_{ab},\hat{\rho}_{ab}] 
        + \kappa_1 \mathcal{D}[\hat{a}-\hat{b}]\hat{\rho}_{ab}
        + \kappa_2 \mathcal{D}[\hat{a}^2-\hat{b}^2]\hat{\rho}_{ab}
        ,
    \label{eq:CascadedMEQ_2}
\end{align}
with
\begin{align}
    \hat{H}_{ab} 
        \equiv \hat{H}_a - \hat{H}_b - \fr{i\kappa_1}{2}(\hat{a}^\dag \hat{b}-h.c.)-\fr{i\kappa_2}{2}(\hat{a}^\dag \hat{a}^\dag \hat{b}^2-h.c.).
        \label{eq:cascaded_ham2}
\end{align}
Again, tracing out cavity $b$ from the above equation recovers the cavity $a$ master equation given in Eq.~(\ref{eq:one_mode}), independent of the choice of $\hat{H}_b$.

The next step is the same as before:  we want to pick $\hat{H}_b$ so that cavity $b$ absorbs all photons emitted by cavity $a$ into {\it either} of the two chiral waveguides.  We thus want a pure steady state $\ket{\psi}$ of the two cavity system that is a dark state of both collective loss operators appearing in Eq.~(\ref{eq:CascadedMEQ_2}).  Fortunately, these dark state conditions are not independent:  having $(\hat{a} - \hat{b}) \ket{\psi} = 0$ as before ensures that the state is dark with respect to emission to either waveguide. 

Finally, there remains the question of how exactly to find the desired $\hat{H}_b$.  As we show in Sec.~\ref{sec:soln_of_general_kerr}, the simple educated guess of taking $\hat{H}_b$ to be the negative of $\hat{H}_a$ still works in the presence of two photon driving and loss, and even for a wider class of problems.

\subsection{Connection to Segal-Bargmann representations}
\label{subsec:SBRepresentation}

A second crucial element in our extension of the CQA method is to combine it with the Segal-Bargmann (SB) representation of single-mode pure-state wavefunctions in terms of holomorphic functions \cite{bargmann_hilbert_1961,Segal1962,Segal1963,bargmann_hilbert_1967}.  This provides an extremely efficient way of solving the complex recursion relations that determine the desired dark state wavefunction $\ket{\psi}$.  More importantly, it is an extremely useful tool for developing physical intuition.  It renders the operation of tracing out the auxiliary cavity $b$ trivial, and allows one to directly obtain the Wigner function of the cavity-$a$ steady state.

\subsubsection{Basics of the representation}

Consider a single bosonic mode in a pure state $\ket{\psi}$ that is written in terms of Fock states $\ket{m}$ as:
\begin{equation}
    \ket{\psi} = \sum_{m=0}^\infty \alpha_m |m\rangle.
\end{equation}
In the SB representation, this state is associated with a holomorphic function $\psi_{\rm SB}(z)$ defined on the complex plane:
\begin{equation}
    \psi_{\rm SB}(z)  = \sum_{m=0}^\infty \fr{\alpha_m}{\sqrt{m!}} z^m.
    \label{eq:SBDefinition}
\end{equation}
The space of these functions forms a Hilbert space that is unitarily equivalent to the original Fock space, with an induced inner product:
\begin{align}
    \langle \psi_{\rm SB} ,\phi_{\rm SB}\rangle_{{\rm SB}} \equiv  
        \fr{1}{\pi}\int_{\mathbb{C}}d^2z\, \psi_{\rm SB}^*(z) \phi_{\rm SB}(z) e^{-|z|^2}.
\end{align}
The SB wavefunction has a direct physical interpretation:  its modulus determines the Husimi $Q$-function of the state $\ket{\psi}$.  Letting $\ket{z}$ denote a coherent state with amplitude $z$, we have
\begin{equation}
    Q(z) \equiv \frac{1}{\pi} \left| \braket{z}{\psi} \right|^2 
    = \frac{1}{\pi} \left| \psi_{\rm SB}(z^*) \right|^2 e^{-|z|^2}. 
\end{equation}
Finally, the canonical $\hat{c}$ and creation $\hat{c}^\dagger$ operators become linear differential operators in the Bargmann space:
$\hat{c} \mapsto \partial / \partial z$, $\hat{c}^\dag \mapsto z$.

 \begin{figure}
     \centering
     \includegraphics[width=0.8\columnwidth]{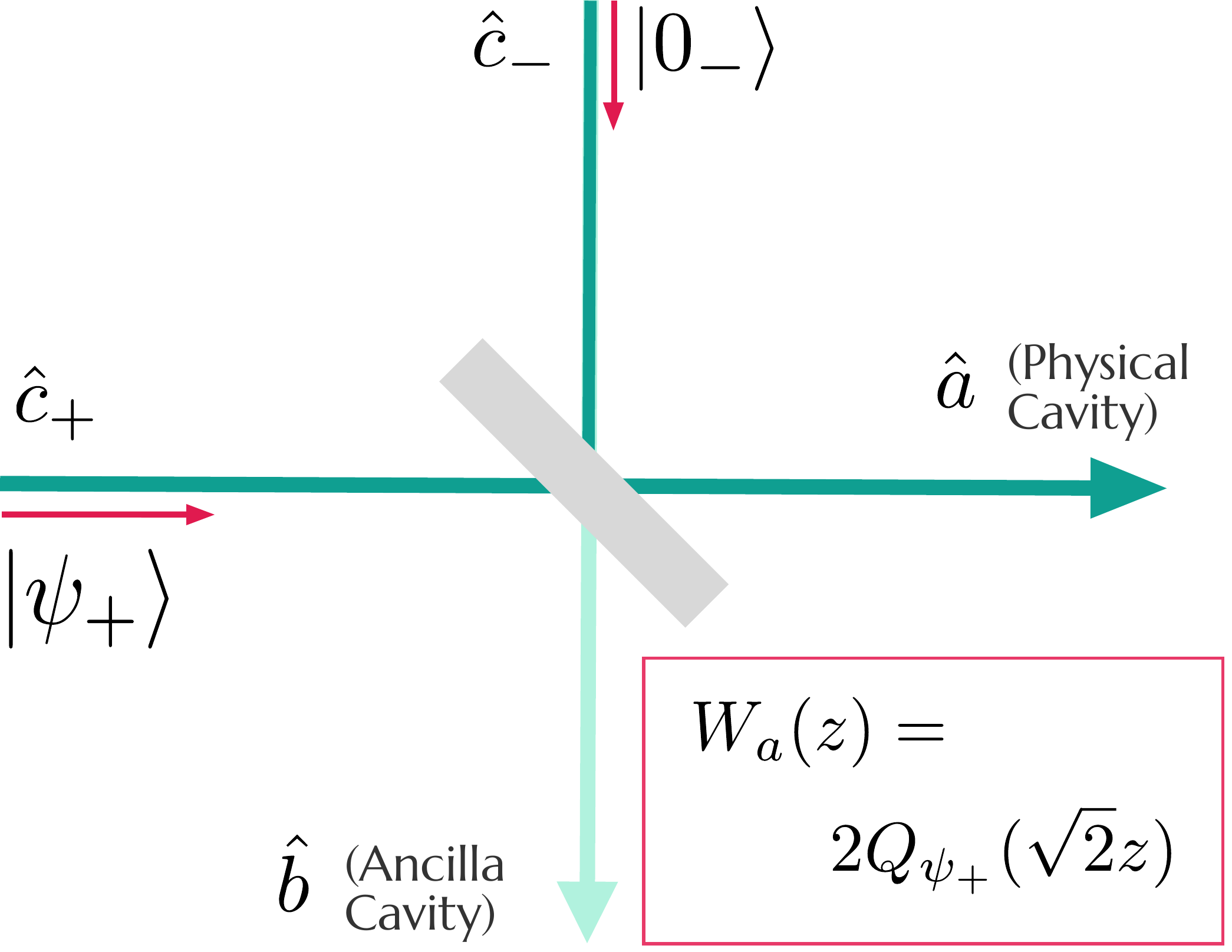}
     \caption{Simple picture of the unique steady state of the generalized driven Kerr resonator with non-zero single-photon loss.  One starts with a pure state, single-mode wavefunction $\ket{\psi_+}$.  This is mixed with vacuum noise at a 50-50 beamsplitter; the output ports represent the final steady state of the physical $a$ cavity and the auxiliary $b$ cavity.
     This operation implies that the cavity-$a$ steady state is $\ket{\psi_+}$ convolved with vacuum noise.  As a result, cavity-$a$'s steady-state Wigner function $W_a(z)$ is equal (up to scaling) to the $Q$ function $Q_{\psi_+}(z)$ of the pure state $\ket{\psi_+}$.  
     }
     \label{fig:beam_splitter}
 \end{figure} 

\subsubsection{Tracing out the auxiliary cavity}

As we will see, the CQA method reduces to finding a single-mode, pure-state wavefunction $\ket{\psi_+}$ for the collective mode $\hat{c}_+ = (\hat{a} + \hat{b})/\sqrt{2}$; the orthogonal mode $\hat{c}_-$ must be in vacuum to have a dark state.  To find the corresponding state of the physical cavity $a$, one transforms from the $\hat{c}_{\pm}$ basis to the $\hat{a} / \hat{b}$ basis, and then traces out the state of the auxiliary cavity $b$.  This operation has a very simple physical interpretation (see Fig.~\ref{fig:beam_splitter}):  it corresponds to mixing the state $\ket{\psi_+}$ with vacuum noise at a 50-50 
beamsplitter, and then discarding one of the output modes. 

At a heuristic level, this operation implies that in phase space, the cavity-$a$ steady state will be equivalent to that of the state $\ket{\psi_+}$ convolved with an extra half-quantum of vacuum noise.  Recall that this is the same transformation that converts a Wigner function into a $Q$-function.  As a result,  
we find a very simple expression for the cavity-$a$ steady state Wigner function.  Letting $\psi_{+,{\rm SB}}(z)$ denote the SB representation of the pure state $\ket{\psi_+}$, we have:
\begin{equation}
    W_a(z) = 2 Q_+(\sqrt{2} z) =  \frac{2}{\pi} \left| \psi_{+,{\rm SB}} \left(\sqrt{2} z^* \right) \right|^2 e^{-2 |z|^2}
    \label{eq:SBWignert}
\end{equation}
We see that the SB ``wavefunction'' $\psi_{+,{\rm SB}}(z)$ has a direct physical interpretation: its modulus determines the cavity-$a$ steady-state Wigner function.  We also see that this Wigner function must necessarily be positive (as it is equivalent to the $Q$-function of the state $\ket{\psi_+}$, and the $Q$ function is always positive).
The above relation follows from the fact that Wigner functions transform in the expected way (i.e.~like classical probability distributions) under a beamsplitter transformation.

A more direct relation results from examination of the cavity-$a$ steady-state $P$-function: since the $P$-function is insensitive to vacuum noise, the output of the beamsplitter in the $P$-representation is simply equal to the $P$-function $P_+(z)$ of the state $|\psi_+\rangle$, rescaled by a factor of $\sqrt{2}$:
\begin{align}
    P_a(z) = 2P_+(\sqrt{2}z).
\end{align}
This immediately implies that the cavity-$a$ $P$ function is generically singular and thus non-positive \cite{cahill_pure_1969}. Thus, in the pathological behavior of the cavity-$a$ $P$-function, we have a tell-tale signature of nonclassicality.

\section{CQA solution of the general driven Kerr cavity} \label{sec:soln_of_general_kerr}

We now use the results of Sec.~\ref{sec:the absorber method} to solve Eq.~(\ref{eq:one_mode}) for a driven Kerr resonator subject to both one and two photon driving, and one and two-photon loss.  This will allow us to reproduce results previously derived using complex-$P$ methods 
\cite{drummond_non-equilibrium_1981,bartolo_exact_2016,minganti_exact_2016,elliott_applications_2016}, but in a manner that allows greater physical intuition.  We are also able to solve an extended model which includes a nonlinear single-photon driving term; this model has not been previously solved.  Our approach yields several new physical insights: the possibility of photon blockade and ``anti-blockade'' phenomena, and the possibility of near quantum-bistability without parity conservation. We focus in this section on the case where there is non-zero single-photon loss ($\kappa_1 \neq 0$), implying the existence of a unique steady state.  In Sec.~\ref{sec:bistability}, we turn to the case where there is no single photon driving or loss; we are able to use the CQA method to provide insights into the bistability in this system, and how this changes from quantum to classical bistability with the addition of a drive detuning.

\subsection{Solution without nonlinear single-photon driving}
\label{subsec:SimpleSoln}

We are interested in the driven-Kerr system described by Eqs.~(\ref{eq:ciuti_ham}) and (\ref{eq:one_mode}) with $\Lambda_3 = 0$ and $\kappa_1 > 0$.  The CQA approach represents this system using the equivalent two-cavity cascaded system in Eqs.~(\ref{eq:CascadedMEQ_2}) and (\ref{eq:cascaded_ham2}).  We seek a pure-state steady-state $\ket{\psi}$ that is necessarily dark with respect to dissipation, meaning that Eq.~(\ref{eq:the absorber methodDarkStateCondition}) is satisfied:  $\sqrt{2} \hat{c}_- \ket{\psi} \equiv (\hat{a}-\hat{b}) \ket{\psi} = 0$.  Our steady state can thus be written as a tensor product of a non-trivial state of the $\hat{c}_+$ collective mode, and a vacuum state for the $\hat{c}_-$ mode:
\begin{equation}
    \ket{\psi} = \ket{\psi_+}\ket{0_-}.
\end{equation}

In order for $\ket{\psi}$ to be a steady state, it also needs to be an eigenstate of the cascaded Hamiltonian $\hat{H}_{ab}$ with energy $E$.  Writing $\hat{H}_{ab}$ in terms of $\hat{c}_{\pm}$, and using the fact that $\hat{c}_{-} \ket{\psi}$ vanishes, the eigenvalue equation becomes
\begin{equation}
    \frac{1}{2}
    \hat{c}^\dagger_{-} \hat{\mathcal{H}}_+ \ket{\psi} = E \ket{\psi},
    \label{eq:NonHermKernel}
\end{equation}
with
\begin{align}
    \hat{\mathcal{H}}_+ & \equiv    
        (K-i\kappa_2)\hat{c}_+^\dag\hat{c}_ +^2 
        -(2\Delta +i\kappa_1)\hat{c}_+ 
        +2\Lambda_2 \hat{c}_+^\dag  +2\sqrt{2}\Lambda_1. 
        \label{eq:NonHermHP}
\end{align}

Our choice of the auxiliary-cavity Hamiltonian $\hat{H}_b$ thus leads to a cascaded Hamiltonian that {\it necessarily} creates an excitation in the $\hat{c}_-$ mode.  It follows that we must have $E=0$.  Having $\ket{\psi}$ be a stationary state then reduces to a single mode problem:
\begin{equation}
    \hat{\mathcal{H}}_+ \ket{\psi_+} = 0,
    \label{eq:the absorber methodKernel}
\end{equation}
i.e.~we need to find a pure state $\ket{\psi_+}$ that is annihilated by the non-Hermitian operator $\hat{\mathcal{H}}_+$.

The seemingly obvious next step is to follow the approach used in Ref.~\cite{stannigel_driven-dissipative_2012}:  express $\ket{\psi_+}$ in the Fock state basis, and turn Eq.~(\ref{eq:the absorber methodKernel}) into a recursion relation for the expansion coefficients $\alpha_j$:
\begin{align}
    \Big[ (K-i\kappa_2)m-(2\Delta +i\kappa_1) \Big]\sqrt{m+1} \alpha_{m+1} 
    \label{eq:Recursion}
    &&\\
         + 2\Lambda_2 \sqrt{m} \alpha_{m-1} 
         +2\sqrt{2}\Lambda_1 \alpha_{m} & =0.\nonumber
\end{align}
In special cases, this reduces to an easily solvable single-term recursion relation: either the case of no two-photon driving $\Lambda_2 = 0$ \cite{stannigel_driven-dissipative_2012}, or the case of no one-photon driving $\Lambda_1 = 0$ \cite{mamaev_entangled_2018}.  In the more general case, the resulting two-term recursion relation is more unwieldy.

A more direct way of getting the desired solution is to use the SB representation $\psi_{+,{\rm SB}}(z)$ of the state $\ket{\psi_+}$.  Eq.~(\ref{eq:the absorber methodKernel}) is then transformed into a second-order ordinary differential equation:
\begin{align}
\Big[ z\pdd{}{z}&-D\pd{}{z}+(\lambda_2 z +\lambda_1)\Big]\psi_{+,{\rm SB}} (z)=0,   \label{eq:complicated_recursion_2}
\end{align}
where 
\begin{align}
        D & = \frac{2 \widetilde{\Delta}}{\widetilde{K}}, ~~~
        \lambda_2 = \fr{2\Lambda_2}{\widetilde{K}},~~~
        \lambda_1 = \fr{2\sqrt{2}\Lambda_1}{\widetilde{K}}.
        \label{eq:original_constants}
\end{align}
Here, $\widetilde{\Delta}\equiv \Delta +i\kappa_1/2$ and $\widetilde{K}\equiv K-i\kappa_2$ are, respectively, effective complex detuning and Kerr nonlinearity parameters.  

Without two-photon driving (i.e.~$\lambda_2 = 0$), Eq.~(\ref{eq:complicated_recursion_2}) is a standard hypergeometric equation.  It has a unique analytic solution:
\begin{align}
    \psi_{+,{\rm SB}}(z)&=Nz^{(D+1)/2}
    J_{-(D+1)}\left(2\sqrt{\lambda_1 z}\right),\label{eq:bessel_soln}
\end{align}
where $J_n(x)$ is a Bessel function and $N$ is a normalization constant.  Using the correspondence between the SB wavefunction and Fock state amplitudes (c.f.~Eq.~(\ref{eq:SBDefinition})), we recover the infinite series result given in Ref.~\cite{stannigel_driven-dissipative_2012}, which in turn corresponds to the classic solution of Ref.~\cite{drummond_quantum_1980}.
The closed form we have here has additional virtues.  Via Eq.~(\ref{eq:SBWignert}), it directly yields a closed form expression for the steady-state Wigner function of the physical cavity $a$; this is in contrast to expressions involving infinite sums that are the usual result of complex-$P$ solutions.  Our expression for this case agrees with that derived earlier (via an alternate method) \cite{kheruntsyan_wigner_1999}.

We turn now to the more interesting case where $\lambda_2 \neq 0$.  Eq.~(\ref{eq:Recursion}) is now a more nontrivial second-order recursion relation.  
The SB representation allows us, however, to simplify the system via non-standard transformations. An example is a ``non-unitary gauge transformation''
\begin{align}
    \psi_{+,\rm SB}(z) \equiv e^{-\theta(z)}\phi(z),\label{eq:translation}
\end{align}
where $\theta(z)$ is the ``gauge potential''. This transformation shifts the differentiation operator by the gradient of $\theta(z)$, $\partial_z \mapsto  \partial_z -\partial_z \theta(z)$.
Here, we try the simplest potential $\theta(z) \equiv \epsilon z$, with $\epsilon$ some constant. Note that as $\theta$ is not purely imaginary, the resulting transformation on the Hilbert space is non-unitary. In the Fock representation, it is equivalent to acting on the state by the exponential of a raising operator: 
\begin{equation}
    \ket{\psi_+} \propto e^{-\epsilon \hat{c}_+^\dagger } \ket{\phi}.
        \label{eq:MappingStates}
\end{equation}

After our transformation, the problematic two-photon driving term is effectively shifted by an amount $\epsilon^2$:
\begin{align}
    \Big[z\pdd{}{z}-&(2\epsilon z+D)\pd{}{z}\nonumber\\
    &+(\lambda_2+\epsilon^2)z +(\lambda_1+\epsilon D)\Big]\phi(z)=0.\label{eq:complicated_recursion_2B}
\end{align}
It can thus be eliminated by choosing $\epsilon$ such that
\begin{equation}
    \epsilon_\pm = \pm i\sqrt{\lambda_2}.
   \label{eq:SBTransformed}
\end{equation}
We will call these non-unitary gauges {\it plus}-gauge and {\it minus}-gauge. Choosing, e.g.~the plus gauge $\epsilon\equiv \epsilon_+$, we see that the gauge-transformed state $\phi(z)$ satisfies Kummer's differential equation (see \cite{brychkov_handbook_nodate}), so that:
\begin{align}
    \phi(z)=N_0\bigg[\,_1F_1\bigg(-\fr{\lambda_1+\epsilon D}{2\epsilon};-D; 2\epsilon z\bigg)\bigg],
    \label{eq:kumm}
\end{align}
where $N_0$ is a normalization factor, and $\,_1F_1(r_1;r_2;z)$ is Kummer's  hypergeometric function, the same special function which appears in the hydrogen atom problem (see, e.g. \cite{sakurai_modern_2017}).  We stress that that the special case where $D$ is a positive integer must be treated specially; this is discussed in Sec.~\ref{sec:Insights}.  Note also that in the $\epsilon \rightarrow 0$ limit, the solution above tends smoothly to the Bessel-function solution in Eq.~(\ref{eq:bessel_soln}).  

The above result combined with Eq.~(\ref{eq:SBWignert}) immediately yields a closed-form expression for the steady-state Wigner function of the physical $a$ cavity of interest: 
\begin{align}
    W_{a, ss}(z) = N 
        |\phi (\sqrt{2}z^* ) |^2\,e^{-2 |z+ \epsilon/\sqrt{2} |^2},\label{eq:closedform}
\end{align}
where $N$ is a normalization constant. Note that if $\phi(z) = 1$, then $W_{a,ss}(z)$ corresponds to a coherent state with amplitude $\alpha = \sqrt{-\lambda_2/2}$.  Thus, a non-unity $\phi(z)$ describes corrections to the dark state being just a simple coherent state.  Note also that if one had chosen the minus gauge in Eq.~(\ref{eq:translation}), one obtains an identical solution (see Appendix \ref{app:AlternateTransform}).

\subsection{Including nonlinear single-photon driving}
We now allow $\Lambda_3 \neq 0$ in Eq.~(\ref{eq:ciuti_ham}).  We are still able to exactly solve for the steady state in this case; unless $\kappa_2 = 0$, it has a qualitatively different form from the $\Lambda_3 = 0$ case.  
The CQA method proceeds as in Sec.~\ref{subsec:SimpleSoln}.  We again write the two-mode dark state as $|\psi\rangle = |\psi_+\rangle |0_-\rangle$, and the eigenvalue equation again reduces to finding the kernel of a non-Hermitian operator $\hat{\mathcal H}_+$:
\begin{align}
    \hat{\mathcal H}_+ = 
        \Big(\widetilde{K}\hat{c}_+^\dag+\sqrt{2}\Lambda_3^* \Big)\hat{c}_+^2 & + \Big(2\sqrt{2}\Lambda_3\hat{c}_+^\dag-2\widetilde{\Delta} \Big)\hat{c}_+\nonumber \\
    & + \left(2\Lambda_2\hat{c}_+^\dag+2\sqrt{2}\Lambda_1 \right).
\end{align}
Comparing against Eq.~(\ref{eq:NonHermHP}), we see that the presence of $\Lambda_3$ creates a term proportional to $\hat{c}_+^2$.  Attempting to solve directly for $\ket{\psi_+}$ in the Fock basis leads a complicated recursion relation, as now we have terms that add a photon $(\propto \hat{c}_+^\dag)$, as well as those that subtract two photons $(\propto \hat{c}_+^2)$. One obtains a third-order recursion, in place of the\ second-order recursion that we had before.

One can nonetheless still solve for the dark state in closed-form.  We first perform a displacement,
\begin{align}
    |\xi_+\rangle =\hat{D}(\alpha_+)|\psi_+\rangle,
    \label{eq:displacement}
\end{align}
where $\alpha_+=\sqrt{2}\Lambda_3/\widetilde{K}^*$, and $\hat{D}(\alpha)\equiv e^{\alpha \hat{c}^\dag_+ -h.c.}$ is the standard displacement operator. We can then remove the two-photon drive by applying a non-unitary gauge transformation (as before), yielding a differential equation which again has a simple solution in terms of Kummer's confluent hypergeometric function:
\begin{align}
    \phi(z)&=N_0\bigg[\,_1F_1
    \bigg(-\fr{\lambda_1+\epsilon_+D}{\epsilon_+-\epsilon_-};-D;(\epsilon_+ -\epsilon_-)z\bigg)\bigg].
    \label{eq:Kumm2}
\end{align}
Here, $\epsilon_\pm$  correspond to the non-unitary gauge choices in which the displaced two-photon drive vanishes (c.f. Eq. (\ref{eq:SBTransformed})):
\begin{align}
    \lambda_2-\lambda_3\epsilon +\epsilon^2&=0\label{eq:SBTransformed2}
\end{align}
To be manifestly consistent with the solution of the driven Kerr cavity without nonlinear coherent driving, we have again written the solution in the {\it plus} gauge. Finally, $\lambda_3,\lambda_2,\lambda_1, D$ are the following general complex constants:
\begin{align}
    D &= \fr{2}{\widetilde{K}}\bigg(\widetilde{\Delta}+\fr{2|\Lambda_3|^2}{\widetilde{K}}\bigg),
    \label{eq:D}\\
    \lambda_1 &= 
        \fr{\sqrt{2}\Lambda_3}{|\widetilde{K}|^2}\bigg(\fr{4|\Lambda_3|^2}{\widetilde{K}}+2\widetilde{\Delta}\bigg)+\fr{2\sqrt{2}}{\widetilde{K}}\bigg(\Lambda_1-\fr{\Lambda_2\Lambda_3^*}{\widetilde{K}}\bigg),
    \label{eq:l1}\\
    \lambda_3&=\fr{2\sqrt{2}\Lambda_3}{\widetilde{K}}\bigg(1-\fr{\widetilde{K}}{\widetilde{K}^*}\bigg),~~\lambda_2=\fr{2\Lambda_3^2}{|\widetilde{K}|^2}\bigg(\fr{\widetilde{K}}{\widetilde{K}^*}-2\bigg)+\fr{2\Lambda_2}{\widetilde{K}}.
    \label{eq:Lambda3}
\end{align}
We have again defined $\tilde{\Delta} = \Delta + i \kappa_1/2$, $\tilde{K} = K - i \kappa_2$.  
For the case where $\Lambda_3 \rightarrow 0$, these parameters revert to those given before Eq. (\ref{eq:original_constants}).  Note that for vanishing two-photon loss, $\widetilde{K}$ is real, and hence Eq.~(\ref{eq:Lambda3}) implies that $\lambda_3 = 0$.  In this case, the cubic drive does not give us anything qualitatively new, as it can be completely eliminated by our displacement transformation.  In contrast, for non-zero $\kappa_2$, cubic driving gives rise to genuinely new phenomena.

As before, the solution above directly determines the steady-state Wigner function of the physical cavity:
\begin{align}
W_{a,ss}(z-\alpha)&=N|\phi(\sqrt{2}z^*)|^2\,e^{-2|z+\epsilon_+/\sqrt{2}|^2}, \label{eq:general_wigner}
\end{align}
where $\alpha \equiv \alpha_+/\sqrt{2}$, and $N$ is a normalization constant. Note that, if $\lambda_3\equiv 0$, then the non-unitary gauge choices in Eq. (\ref{eq:SBTransformed2}) satisfy $\epsilon_+ = - \epsilon_-$, and so $\epsilon_+-\epsilon_- \to 2\epsilon_+$, and we recover the standard solution Eq. (\ref{eq:closedform}).

\section{Steady-state phase diagram of the generalized driven Kerr resonator}
\label{sec:Insights}
We now use our exact solutions in Eqs.~(\ref{eq:kumm}) and (\ref{eq:Kumm2}) to explore the parameter dependence of the steady state of our generalized driven-dissipative Kerr resonator. The steady-state is largely controlled by just two dimensionless parameters $r_1, r_2$.  For the usual case $\Lambda_3 = 0$ (no three photon drive), these are:
\begin{align}
r_1 & \equiv  \fr{\lambda_1+\epsilon D}{2\epsilon}
 =\fr{\Delta+i \frac{\kappa_1}{2}}{K-i\kappa_2} 
       - \frac{ i\Lambda_1 }{\sqrt{\Lambda_2 (K-i\kappa_2)} }
       \label{eq:r1Definition} \\
r_2& \equiv D
=\fr{2 \Delta+i\kappa_1}{K-i\kappa_2}.
        \label{eq:r2Definition} 
\end{align}
The various drive amplitudes $\Lambda_j$ enter only through $r_1$; in contrast, $r_2$ is a generalized detuning parameter which is independent of drive amplitudes.  With a non-zero $\Lambda_3$, one has 
$r_1 = (\lambda_1+\epsilon_+D)/(\epsilon_+-\epsilon_-)$, $r_2 = D$, where 
$\lambda_1, D, \epsilon_\pm $ are defined in Eqs.~(\ref{eq:SBTransformed2})-(\ref{eq:Lambda3}).

\begin{figure}
     \centering
     \includegraphics[width=0.95\columnwidth]{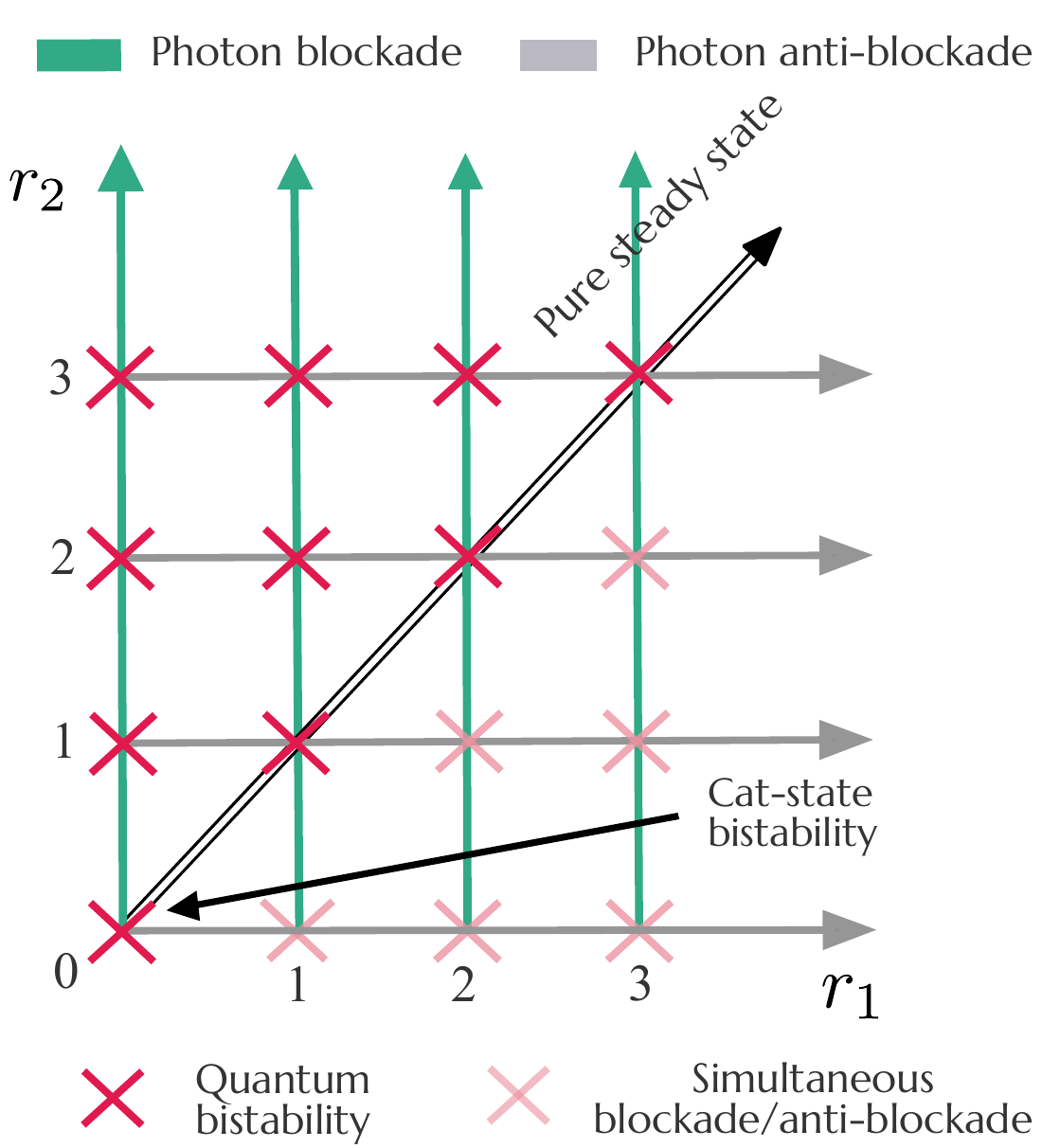}
     \caption{Steady-state phase diagram for the generalized, driven-dissipative Kerr resonator. $r_2$ is a dimensionless detuning parameter, whereas $r_1$ is a drive-dependent dimensionless parameter; both are defined in Eqs.~(\ref{eq:r1Definition})-(\ref{eq:r2Definition}).  The phase diagram indicates parameter choices that lead to unusual steady states (as discussed in the main text).  }
     \label{fig:phase_diagram}
 \end{figure}

As we now show, the steady state exhibits remarkable properties whenever system parameters are tuned to make one or both of $r_1,r_2$ be non-negative integers (see Fig.~\ref{fig:phase_diagram}).  At these points in parameter space, the solution can exhibit generalized forms of photon blockade and anti-blockade, as well as new kinds of bistability.  This latter result generalizes the previously studied cat-state bistability that occurs when $\Lambda_1 = \Delta = \kappa_1 = 0$ (i.e.~$r_1 = r_2 = 0$) \cite{mirrahimi_dynamically_2014}.  We stress that all of these features have clear observable signatures, and are quantum in nature.  In what follows, we focus primarily on the standard case $\Lambda_3 = 0$.  We also highlight the fact that with the addition of a nonlinear coherent drive, the observable consequences of the photon blockade and anti-blockade phenomena can be made even more dramatic.

\subsection{Basic intuition}

Recall that the steady state is determined by a single-mode pure-state  $\ket{\psi_+}$ 
(c.f.~Fig.~(\ref{fig:beam_splitter})), and that further, this state is related to a simpler state $\ket{\phi}$ via a ``non-unitary gauge transformation'' (c.f.~Eq.~(\ref{eq:MappingStates})). We could always expand the transformed state $\ket{\phi}$ in the Fock basis as:  
\begin{equation}
    \ket{\phi} = 
        \sum_{m=0}^{\infty} \beta_m \ket{m}_+.
        \label{eq:EffectiveState}
\end{equation}
Defining the scaled Fock state amplitudes (c.f.~Eq.~(\ref{eq:EffectiveState}))
\begin{equation}
    c_m = \beta_m \frac{\sqrt{m!}}
    { \left(2\epsilon\right)^{m}}
\end{equation}
the ODE defining the gauge-transformed state $\ket{\phi}$ in Eq.~(\ref{eq:kumm}) is equivalent to the simple recursion relation ($m \geq 0$)
\begin{equation}
    (m-r_2) c_{m+1} = (m-r_1) c_m
    \label{eq:EffectiveRecurrence}
\end{equation}
with $r_1, r_2$ defined in Eqs.~(\ref{eq:r1Definition}),(\ref{eq:r2Definition}). 
The significance of $r_1, r_2$ being positive integers is now clear:  in this case, there is the possibility of the recursion relation terminating (i.e.~vanishing for certain values of $m$).  This termination corresponds to a kind of quantum interference effect, and will be at the heart of the new blockade, anti-blockade and bistability phenomena we describe.  

Note that we can directly go from the Fock state structure of $\ket{\phi}$ to the SB wavefunction of the desired, untransformed state $\ket{\psi_+}$.  For $\Lambda_3 = 0$ the SB wavefunction of $\ket{\psi_+}$ is 
\begin{align}
    \psi_{+,{\rm SB}}(z)
        \propto  e^{-\epsilon z} \sum_{m=0}^\infty  c_m \fr{\left(2\epsilon z\right)^m}{m!},\label{eq:dkstate_recent}\\
\nonumber
\end{align}
with $\epsilon = i\sqrt{\lambda_2}$.  For the more general case with non-zero $\Lambda_3$, up to a displacement, we have:
\begin{align}
    \psi_{+, \rm SB}(z)
        \propto e^{-\epsilon_+z} \sum_{m=0}^{\infty} c_m\fr{\{(\epsilon_+-\epsilon_-) z\}^m}{m!}
    \label{eq:SBDarkStateGeneral}
\end{align}
where $\epsilon_\pm$ are defined in Eq.~(\ref{eq:SBTransformed2}).  Recall that these SB wavefunctions directly determine the steady-state Wigner function of the physical cavity via Eq.~(\ref{eq:SBWignert}).

\subsection{Pure unique steady states: $r_1 = r_2$}

The first surprising phenomena we describe is the emergence of unique pure steady states even with nonlinearity.  In general, the combination of dissipation and nonlinearity leads us to anticipate impure cavity-$a$ steady states.  Surprisingly, there are a range of parameters where the unique steady state of cavity $a$ is a pure coherent state (as would be expected from a damped, linearly-driven, linear cavity).  This occurs when parameters are chosen such that $r_1 = r_2$ (without either being a positive integer).  In terms of physical parameters, and for $\Lambda_3 = 0$, this requires tuning the one and two photon drives $\Lambda_1, \Lambda_2$ so that:
\begin{equation}
        - \frac{ \Lambda_1 }{\sqrt{ -\Lambda_2 (K-i\kappa_2)} } = 
        \fr{\Delta+i \frac{\kappa_1}{2}}{K-i\kappa_2} 
\end{equation}

For this parameter tuning, Eq.~(\ref{eq:EffectiveRecurrence}) implies that all the scaled Fock state amplitudes $c_m$ are identical.  This in turn implies from Eq.~(\ref{eq:dkstate_recent}) and (\ref{eq:SBDefinition}) that the state $\ket{\psi_+}$ is a coherent state with amplitude 
$\gamma =i \sqrt{ 2 \Lambda_2 / (K - i \kappa_2)} = i \sqrt{\lambda_2}$.  As sending coherent states through a beamsplitter also generates coherent states at the output, this also implies that the cavity-$a$ steady state is a simple, pure coherent state of amplitude $ \gamma / \sqrt{2}$.  This follows directly from Eq.~(\ref{eq:dkstate_recent}) and the general expression in Eq.~(\ref{eq:SBWignert}) for the steady-state cavity-$a$ Wigner function.  Note that this steady-state coherent state amplitude is consistent with the semiclassical cavity-$a$ equations of motion.

\subsection{Higher-order photon blockade: $r_1 = n_0$}
\label{subsec:PhotonBlockade}

Surprising effects also occur when drives and detuning are chosen so that $r_1 = n_0$, where $n_0$ is a non-negative integer.  The recursion relation in Eq.~(\ref{eq:EffectiveRecurrence}) now terminates at $m = n_0$: Fock state amplitudes $c_m$ vanish for all $m \geq n_0 +1$.  This is an example of a generalized strong photon-blockade phenomena:  the gauge-transformed steady-state $\ket{\phi}$  has strictly zero probability to have more than $n_0$ photons. Unlike standard  photon blockade \cite{imamoglu_strongly_1997}, the mechanism here does not require infinitely strong nonlinearity.  Also, unlike the so-called ``unconventional'' photon blockade \cite{liew_single_2010,bamba_origin_2011,lemonde_antibunching_2014}, the blockade here is complete:  there is strictly no probability to have more than $n_0$ photons in the state. 

\begin{figure}
     \centering
     \includegraphics[width=0.999\columnwidth]{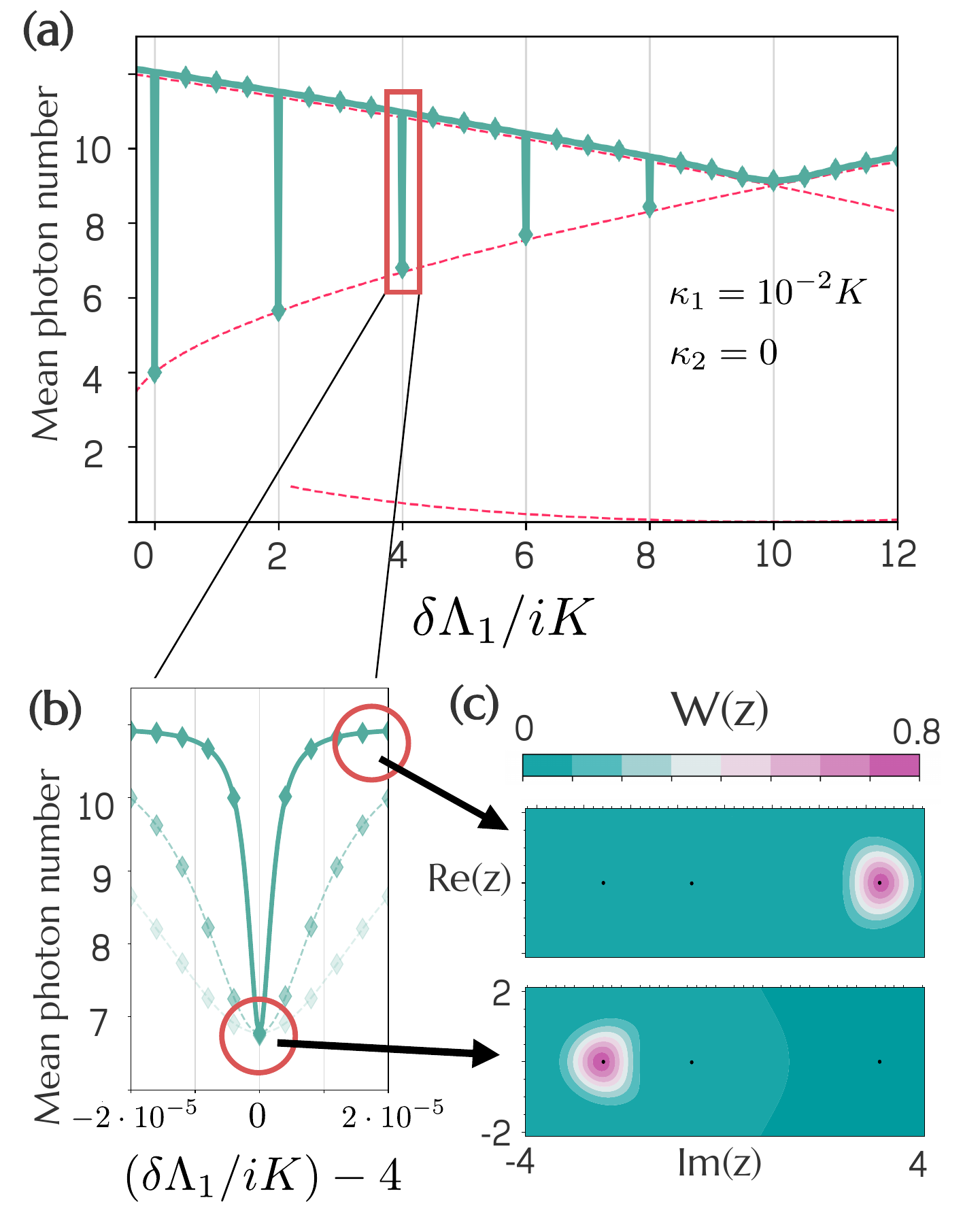}
     \caption{
     {\bf Generalized Photon Blockade.}
     (a) Mean steady-state cavity-$a$ photon number
     as a function of single photon drive 
     $\delta \Lambda_1 \equiv \Lambda_1-\Lambda_1^{(0)}$, where the offset $\Lambda_1^{(0)}=(0.01-10i)\cdot K$ is determined from  
     Eq.~(\ref{eq:regularly_spaced}) and our choice of system parameters.   The periodic, sharp drop in photon number corresponds to a generalized photon blockade phenomena, which occurs whenever the parameter $r_1$ (c.f.~Eq.~(\ref{eq:r1Definition}))is a non-negative integer.  Solid lines:  analytic exact solution, diamonds: master equation numerics.    Photon numbers associated with the semiclassical stationary stable amplitudes are also plotted (dashed red lines).
     (b) Zoom-in of one of the blockade anti-resonances. Loss values are $\kappa_1 = K/100, K/20,$ and $K/10$, with more faded green corresponding to greater loss. (c) Steady-state Wigner function, for two choices of $\delta \Lambda_1$ corresponding to being either at (off) a blockaded parameter value; black dots indicate the three semiclassical amplitudes that exist for these parameters.  For all results, $\Delta = 5K$, $\Lambda_2=4K$ , $\kappa_1=10^{-2}K$, and $\Lambda_3 = \kappa_2=0$.  By using nonlinear coherent driving $\Lambda_3$, this blockade phenomenon can be made sharp (i.e.~there is a sharp cutoff in the photon number distribution).
     }
     \label{fig:blockades}
 \end{figure}

While the ``gauge-transformed'' state $\ket{\phi}$ exhibits blockade, physical phenomena is controlled by the untransformed state $\ket{\psi_+}$.    
 Eq.~(\ref{eq:MappingStates}) shows that this state is a ``smeared'' version of the blockaded state. Despite this, the physical cavity 
$a$ steady state still shows a pronounced suppressed photon population whenever the parameter $r_1$ is tuned to an integer.  This 
blockade-induced suppression can be observed by considering how the steady state changes as a function of the single photon drive amplitude $\Lambda_1$ (as this tunes $r_1$ but not $r_2$).  From Eq.~(\ref{eq:r1Definition}), one sees that blockade occurs periodically as a function of $\Lambda_1$, with the $n$th-order blockade occuring when 
\begin{align}
    \Lambda_1 \equiv 
    \Lambda_1^{(0)}-in \sqrt{ [K - i \kappa_2] \Lambda_2 },
    \label{eq:regularly_spaced}
\end{align}
where 
$\Lambda_1^{(0)} = -i
(\Delta + i \kappa_1 / 2) \sqrt{\Lambda_2}(K - i \kappa_2)^{-3/2}$ is a constant offset.  
Note that achieving a blockade requires tuning both the phase and magnitude of the single photon drive amplitude $\Lambda_1$.

Fig.~\ref{fig:blockades} shows representative results for $\kappa_2 = \Lambda_3 = 0$:  the average cavity photon number shows a sharp suppression whenever $\Lambda_1$ is tuned to make $r_1$ a positive integer. 
Note the remarkable fact that the width of these blockade suppressions (as a function of $\Lambda_1$) are much smaller than $\kappa_1$. We stress that in the main plot Fig.~\ref{fig:blockades}(a), it is {\it only} the single photon drive that is being tuned; all other parameters are held fixed. 

Fig.~\ref{fig:blockades}a also plots the photon number associated with each stable, stationary semi-classical amplitude (obtained by solving the classical, noise-free equation of motion).  These semiclassical solutions do not exhibit any sharp behaviour as a function of $\Lambda_1$.  The sharp behaviour of the quantum steady state that occurs when $r_1$ is tuned to a positive integer corresponds to the quantum steady state solution suddenly switching (as a function of $\Lambda_1$) from being localized near the high amplitude classical solution to being localized near a low amplitude classical solution (see Fig.~\ref{fig:blockades}(c)).  The physics here is thus intimately related to physics of quantum activiation and quantum tunneling \cite{marthaler_switching_2006, dykman_critical_2007}, e.g.~the dynamical switching between different semiclassical solutions. For more detailed discussion of semiclassical switching behavior in steady-states of Kerr resonators, see \cite{bartolo_exact_2016,minganti_spectral_2018}. We stress that the behaviour here cannot be understood in terms of the metapotential $M(x,y)$ often used in studies of nonlinear cavities (see, e.g., \cite{dykman_periodically_2012,puri_engineering_2017}).  The metapotential is simply the classical Hamiltonian viewed as a function of the canonical quadratures $x,y$.  It is a completely smooth function of parameters.  For $\Lambda_1 = 0$, it has two degenerate extrema, correspondig to the two stable classical steady states.  Adding an approximately purely imaginary $\Lambda_1$ (as we do in Fig. 4) tilts this metapotential, but does not break the degneracy between the classical solutions.  Hence, this does not provide any insight into why the quantum steady state localizes around one classical amplitude versus another.

\begin{figure}
     \centering
     \includegraphics[width=0.999\columnwidth]{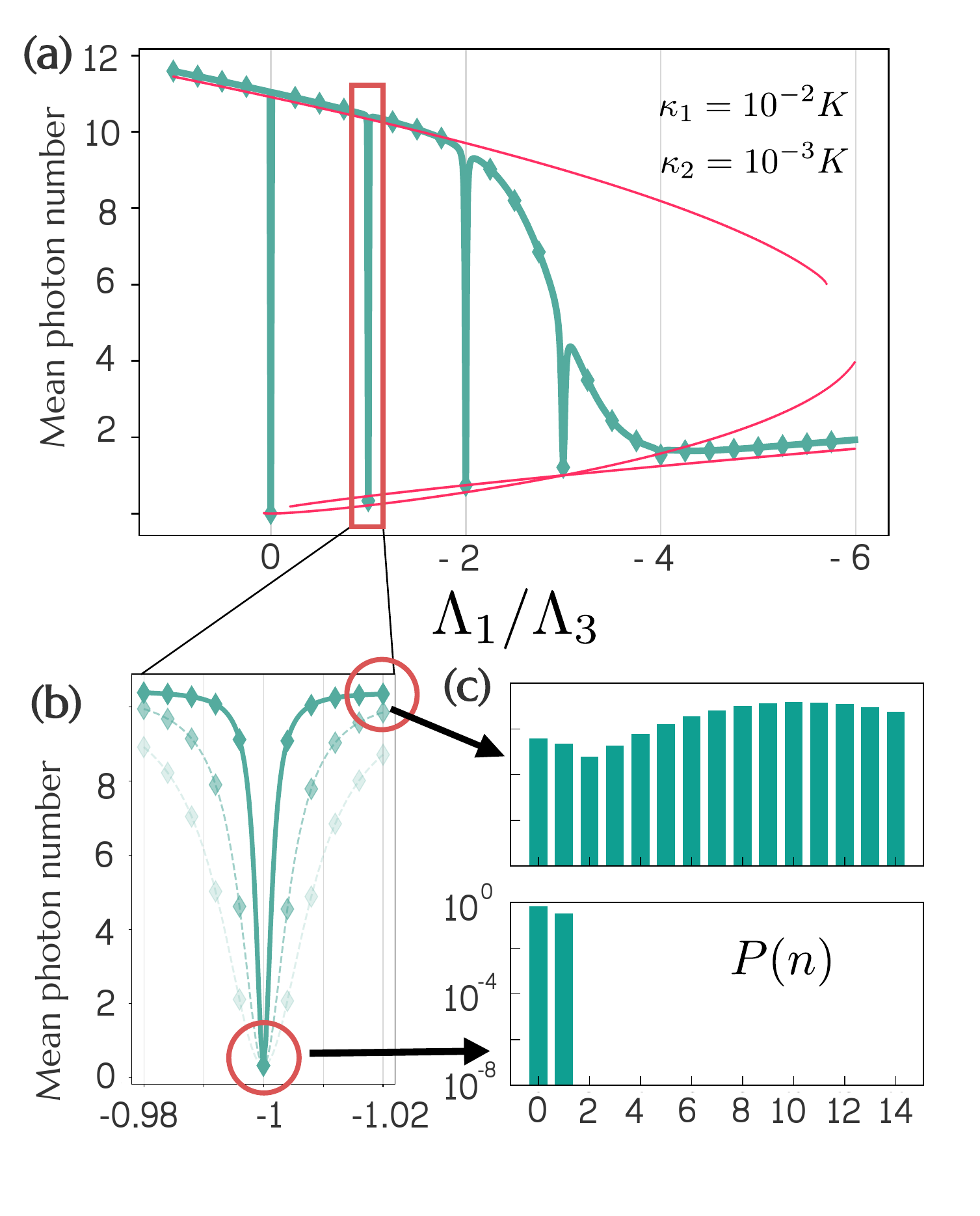}
     \caption{
     {\bf Exact photon blockade using a nonlinear single-photon drive.}
     (a) Mean steady-state cavity-$a$ photon number
     as a function of single photon drive 
     $\Lambda_1$.  The periodic, sharp drop in photon number corresponds to a generalized photon blockade phenomena, which occurs whenever the parameter $r_1$ (c.f.~Eq.~(\ref{eq:r1Definition}))is a non-negative integer.  Solid lines:  analytic exact solution, diamonds: master equation numerics.    Photon numbers associated with the semiclassical stable amplitudes are also plotted (red lines).
     (b) Zoom-in of one of the blockade anti-resonances. Loss values are $\kappa_1 = K/100, K/20,$ and $K/10$, with more faded green corresponding to greater loss. (c) Steady-state photon statistics, for two choices of $\Lambda_1$ corresponding to being either at (off) a blockaded parameter value.  For all results, $\Delta = K$, $\Lambda_2=0$ , $\kappa_1=K/100$, $\kappa_2=K/1000$, and $\Lambda_3 =K$.
     }
     \label{fig:blockades_exact}
 \end{figure}

Finally, we note that when $r_1 = n_0$, the SB wavefunction for the dark state $\ket{\psi_+}$ (which directly determines the cavity-$a$ Wigner function) reduces to an associated Laguerre polynomial $L^{(\alpha)}_m(z)$:
\begin{align}
    \psi_{+,{\rm SB}}(z) \underset{r_1\to n_0}
    \propto 
    e^{-\epsilon z}L_{n_0}^{(1-D)}
    \left(2\epsilon z \right),
\end{align}
where for $\Lambda_3 = 0$, we have $\epsilon = i\sqrt{\lambda_2}$.

\subsection{Sharp photon blockade with weak nonlinearities}

We now show that the generalized photon blockade phenomena is most striking and intuitive in the case where there is no two photon driving, but only linear and nonlinear one photon driving $\Lambda_1, \Lambda_3 \neq 0$.  Recall first our most general solution which includes a non-zero $\Lambda_2$.  In terms of the displacement parameter $\alpha_+ = \sqrt{2}\Lambda_3/\widetilde{K}^*$, the desired steady-state, pure state of the $+$ mode is given by:
\begin{align}
    |\psi_+\rangle = \hat{D}(\alpha_+)e^{-\epsilon_+ \hat{c}_+^\dag}|\phi\rangle
\end{align}
where $|\phi\rangle$ is the core state defined in Eq. \eqref{eq:EffectiveState}. The photon blockade phenomenon is best intuitively understood when there is nonlinear one-photon driving, but no two-photon driving. In the limit that $\Lambda_2$ vanishes, $\alpha_+\to -\epsilon_+$, meaning that, after direct application of the Baker-Campbell Hausdorff identity, the displacement transformation partially cancels the exponential factor:
\begin{align}
    \psi_{+,\rm SB}(z)\underset{\Lambda_2\to 0}{\sim} \phi_{\rm SB}(z+\alpha_+)
\end{align}
where $\phi_{\rm SB}$ is the Segal-Bargmann representation of the core state defined in Eq. \eqref{eq:EffectiveState}. Therefore, we see that the limit $\Lambda_2\to 0$ is physically important because it removes the smearing factors that spoil the bare physics contained in the state $|\phi\rangle$.

Further, in the limit of vanishing $\Lambda_2$
\begin{align}
    r_1\underset{\Lambda_2\to 0}{\sim}-\Lambda_1/\Lambda_3.
\end{align}
It thus follows that a sharp photon blockade occurs in the cavity $a$ steady state each time the ratio of nonlinear to linear one-photon driving is a negative integer (representative results are shown in Figure \ref{fig:blockades_exact}).

In this limit, it is easy to understand the origin of the photon blockade phenomena as the result of destructive interference between linear and nonlinear one-photon driving. The Hamiltonian in this case is
\begin{align}
    \hat{H}&=\frac{K}{2}\hat{a}^\dag \hat{a}^\dag \hat{a}\hat{a} -\Delta \hat{a}^\dag \hat{a}\nonumber\\
    &+\Lambda_3(\hat{n}-r_1)\hat{a}^\dag +\Lambda_3^*\hat{a}(\hat{n}-r_1^*)\label{eq:blockaded_ham}
\end{align}
When $r_1$ is tuned to a non-negative integer $n_0$, the Hamiltonian has strictly no matrix elements connecting Fock states with photon number $n_0$ or less to states with photon number $n_0+1$ or greater.  The result is that the system becomes ``trapped'' in the subspace of states having $n_0$ or less photons.   

As this mechanism for photon blockade depends on matrix elements and not energy detunings, it is effective even in regimes where dissipation is much stronger than nonlinearity:  while dissipation can smear out energies, it does not smear out matrix elements, meaning that the interference preventing excitation of the $n_0+1$ Fock state is robust.  To see this explicitly, consider the simplest case $n_0 = 1$, where the system gets stuck in a subspace with at most one photon.  We also consider for simplicity a system where the only nonlinearity is the nonlinear drive $\Lambda_3$ (i.e. $K = \kappa_2 = 0$), and where there is no drive detuning $\Delta$.  In this case, the steady state depends only on a single dimenionless parameter $\Lambda\equiv  \Lambda_3/\kappa_1$, and can be found using elementary  means.  This state only involves the vacuum state $\ket{0}$ and $n=1$ Fock state $\ket{1}$ and is given by:

\begin{align}
    \hat{\rho}_\text{ss} & = 
        \frac{
            (4 \Lambda^2+1) \ket{0}\bra{0} + 
            4 \Lambda^2 \ket{1}\bra{1} + 
            2i \Lambda \left(\ket{1}\bra{0} - h.c. \right) 
                }
    {8\Lambda^2+1}
\label{eq:simple_blockaded_state}
\end{align}

In the limit of weak nonlinearity $\Lambda \rightarrow 0$, the blockade of course still remains sharp:  there is still zero probability for the state to have $2$ or more photons, even though the one-photon probability is $\Lambda^2 / 2 \ll 1$.  We stress that this mechanism is completely distinct from the so-called unconvential photon blockade \cite{liew_single_2010,lemonde_antibunching_2014}, which also only requires weak nonlinearities, but which is restricted to Gaussian states, and which does not produce a sharp blockade (e.g.~there is non-zero probability to have more than one photon).   

While at first glance the nonlinear one-photon drive term may seem quite exotic, it is within reach of experiment.  In Appendix \ref{app:cQED}, we show how this driving term could be realized in circuit QED using the recently developed SNAIL architecture \cite{sivak_kerr_2019}.  Generalized photon blockade may have applications in quantum information science settings where nonlinearity is a limited resource.

\subsection{Photon anti-blockade: $r_2 = m_0$}

Tuning the parameter $r_2$ to be an integer $m_0$ in the recurrence relation Eq.~(\ref{eq:EffectiveRecurrence}) also results in 
unusual behaviour of our dark states.  For zero-dissipation, $r_2 = m_0$ is simply the condition for the Fock states $n=0$ and $n= m_0$ of our physical $a$ cavity to be degenerate in the absence of any driving (i.e.~the detuning and Kerr terms cancel out) \cite{bartolo_exact_2016}. Such resonances are analogous to multi-photon resonances that are used to directly drive transmon qubits from the ground state to the $n$th excited state (as a transmon can also be approximately modelled as a Kerr resonator). Our exact solution shows that this resonance condition has strong consequences even with dissipation and drive.  When $r_2 = m_0$, the only solution to the recurrence relation has the coefficients $c_1$ through $c_{m_0}$ be exactly zero.  This implies that the gauge-transformed dark state in Eq.~(\ref{eq:EffectiveState}) will have strictly zero probability to have a photon number equal to $m_0$ or smaller (while higher Fock states will be occupied).  We call this phenomenon a photon ``anti-blockade''.  As with the photon-blockade phenomena, this will also have implications for our physical cavity, via Eq.~(\ref{eq:MappingStates}).

\begin{figure}
    \centering
    \includegraphics[width=0.95\columnwidth]{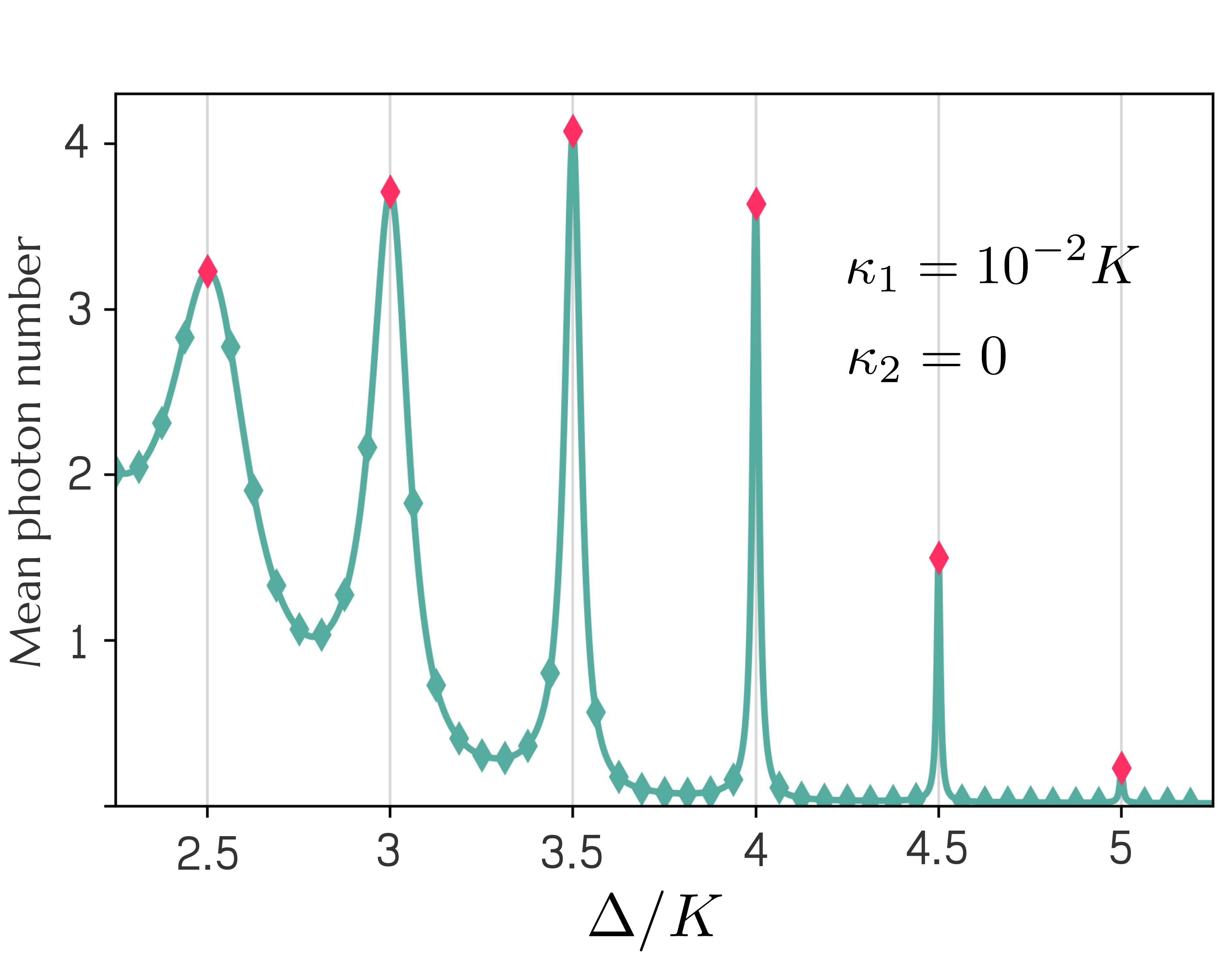}
    \caption{
  {\bf Photon anti-blockade.}
    Average cavity-$a$ steady-state photon number as a function of drive detuning $\Delta$, with drive amplitudes fixed at $\Lambda_1=\Lambda_2=K/2$, $\Lambda_3 = 0$.  Resonances here correspond to having tuned the parameter $r_2$ (c.f.~Eq.~(\ref{eq:r2Definition})) to be near a non-negative integer.  Other parameters are $\kappa_1 = 0.01K$, $\kappa_2 = 0$.
    }
    \label{fig:blockade}
\end{figure}

For any non-zero amount of dissipation, it is clear from Eq.~(\ref{eq:r2Definition}) that we can never have $r_2$ exactly be a positive integer (as $\kappa_1, \kappa_2 \geq 0$).  This remains true even in the presence of a nonlinear coherent drive, where $r_2 = D$, with $D$ given by Eq.~(\ref{eq:D}).
Nonetheless, for weak dissipation (namely $\kappa_1 \ll \Delta$, $\kappa_2 \ll K$), one can still tune $r_2$ to be extremely close to an integer.  In this regime, one still has strong signatures of the anti-blockade behaviour.  For the physical cavity $a$, this translates into a kind of resonant enhancement of photon number and skewed photon number statistics. Representative behaviour is shown in  Fig.~\ref{fig:blockade}.  Note that this resonance phenomenon was observed in Ref.~\cite{bartolo_exact_2016}, though connections to photon statistics and the properties of the analytic steady-state solution were not discussed.


\subsection{Generalized bistability: $(r_1,r_2) = (n_1, n_2)$}
\label{subsec:Bistability}

Having understood photon blockade and anti-blockade phenomena, the natural remaining case is when both these phenomena coexist.  This occurs when parameters are chosen so that
$(r_1,r_2) = (n_1, n_2)$, 
where $n_2 \geq n_1$ are both non-negative integers. Eq.~(\ref{eq:EffectiveRecurrence}) then yields  {\it both} a photon-blockaded solution, and a distinct anti-blockaded solution.  These correspond to two distinct dark states of the $+$ mode, described respectively by SB wavefunctions:
\begin{align}
    \psi_{1,{\rm SB}}(z)&=\fr{e^{-\epsilon z}}{N_1^{1/2}} 
        \sum_{m=0}^{r_1} ~~ \fr{c_m(2\epsilon z)^m}{m!},\label{eq:bsphysics1}\\
    \psi_{2,{\rm SB}}(z)&=
        \fr{e^{-\epsilon z}}{N_2^{1/2}}
        \sum_{m=r_2+1}^\infty  \fr{c_m(2 \epsilon z)^m}{m!}.
    \label{eq:bsphysics2}
\end{align}
Any linear combination of these solutions is also a dark steady-state.  We refer to this situation as ``quantum bistability'': the extended, two-cavity cascaded system in 
Fig.~\ref{fig:virtual_cavity_intro}(b) has an infinite number of steady states, corresponding to
any superposition state of the form:
\begin{equation}
    \ket{\tilde{\psi}[a_1,a_2]} = a_1 \ket{\psi_{1}}_+ \ket{0}_{-} + a_2 \ket{\psi_{2}}_+ \ket{0}_-
\end{equation}
This steady-state structure is conventionally referred to as a (two-dimensional) decoherence-free subspace. This also implies multi-stability for the physical $a$ cavity, which exhibits a two-parameter continuous family of steady states,  
\begin{equation}
    \hat{\rho}_{a,{\rm ss}} = \textrm{tr}_b \left[\ket{\tilde{\psi}[a_1,a_2]} \bra{\tilde{\psi}[a_1,a_2]} \right].
    \label{eq:BistableSteadyState}
\end{equation}
The upshot is that the generalized driven-dissipative Kerr cavity has a multitude of distinct parameter points that yield multi-stability, despite any obvious symmetry.

Unfortunately, we have the same issue as with the anti-blockade phenomena:  non-zero dissipation makes it impossible to exactly tune to 
bistable parameter values {\it except} for the case $n_1=n_2=0$. This is because the constraint of having one or both of $\kappa_1,\kappa_2$ be positive implies $r_2$ cannot be exactly equal to a positive integer (c.f.~Eq.~(\ref{eq:r2Definition})).  The only exactly-achievable bistable point is the  case $n_1 = n_2 = 0$, which can be reached if $\kappa_1 = 0, \kappa_2 > 0$.  This parameter point corresponds to the 
well-studied cat-state bistability in a two-photon driven Kerr resonator \cite{wolinsky_quantum_1988}.

Despite these caveats, the new bistable points are physically relevant:  for weak dissipation, one can come arbitrarily close to them in parameter space, with striking observable consequences for the steady state.  We explore this further in the next section.  We also discuss in Appendix \ref{app:ExactBistability} how one can {\it exactly} achieve the physics of these new bistable points using a non-cascaded version of the two
cavity setup depicted in Fig.~\ref{fig:virtual_cavity_intro}(b).

\subsection{Simultaneous/coexisting blockade and anti-blockade}
What if $r_1,r_2$ are both non-negative integers, and $r_2 < r_1$? In this case, neither the photon-blockaded {\it nor} the photon-resonant solution is permitted. Instead, a {\it medium}-photon number solution exists, and serves as the unique dark state.  For $\Lambda_3 = 0$, we have
\begin{align}
\psi_{+, \rm SB}(z)=\fr{e^{-\epsilon z}}{N^{1/2}}\sum_{r_2+1}^{r_1}c_m\fr{(2\epsilon z)^m}{m!}
\end{align}
with $\epsilon \equiv i\sqrt{\lambda_2}$.  Without the exponential prefactor, this state would exhibit both photon blockade and anti-blockade (i.e.~ its photon number distribution would be cut-off at small and large photon numbers).

 \section{Consequences of new quantum bistable points}
 \label{sec:nearby} 

As discussed in the previous section, there are an infinite number of points in parameter space where our generalized driven-dissipative Kerr resonator is {\it almost} quantum bistable
(c.f.~Fig.~\ref{fig:phase_diagram}). With non-zero one-photon loss, one cannot exactly achieve the required parameter tuning for bistability, but one can come arbitrarily close to a given bistable parameter point.  In this section, we explore the physical consequences of this near-bistability.  We show that there is an extremely strong sensitivity to small parameter changes when one is in this near-bistable regime, and that the unique steady state can be understood as  ``picking-out'' a unique state from the bistable manifold in Eq.~(\ref{eq:BistableSteadyState}).

Suppose we chose parameters that result in $(r_1,r_2)$ being close to integers $(n_1,n_2)$:
\begin{equation}
    r_1 = n_1 + \delta r_1 ,
    \hspace{1 cm}
    r_2 = n_2 + \delta r_2 ,
\end{equation}
These small deviations kill the bistability.  However, for small $\delta r_j$ the resulting pure steady state of the $+$ mode is a particular linear combination of the states $\phi_j(z)$ that span the bistable manifold at $\delta r_j = 0$.  Moreover, the precise form of this combination is {\it extremely} sensitive to parameter variations. 

For example, consider the simple case where the unperturbed recursion parameters are $(r_1,r_2) = (n,n)$.  In this case, the recursion relation Eq. (\ref{eq:EffectiveRecurrence}) simplifies to
\begin{align}
c_{m+1} = \fr{(m-n)-\delta r_1}{(m-n)-\delta r_2} c_m.\label{eq:simple_case}
\end{align}
In the regime that $\delta r_1,\delta r_2\ll 1$, we can see that the ratio $c_{m+1}/c_m$ is essentially $1$, except for the ratio $c_{n+1}/c_n=\delta r_1/\delta r_2$.  Therefore, as $\delta r_1,\delta r_2\to 0$, the unique steady state solution (i.e.~solution to the recursion relation) has the limiting form
\begin{align}
    \psi_\text{+,SB}(z)\underset{\delta r_1,\delta r_2\to 0}{\sim}\psi_\text{1,SB}(z) + \frac{\delta r_1}{\delta r_2}\psi_\text{2,SB}(z)\label{eq:bistable_SB}
\end{align}
as a superposition of the bistable solutions given in Eqs.~(\ref{eq:bsphysics1}),(\ref{eq:bsphysics2}). Note that in writing this equation, we must pick the overall phase of $\psi_{2,{\rm SB}}$ such that the ratio between $c_{n+1}$ (appearing in $\psi_\text{2,SB}$) and $c_n$ (appearing in $\psi_\text{1,SB}$) is precisely $\delta r_1/\delta r_2$. \\

As a result, the unique steady state Wigner function of the physical $a$ cavity will be:
\begin{align}
 W_{a,ss}(z) 
 \simeq  
    \frac{e^{-2 \left|z \right |^2}}{N}
 \bigg| \psi_{1,{\rm SB}}(\sqrt{2}z^*)+\fr{\delta r_1}{\delta r_2}\,\cdot 
 \psi_{2,{\rm SB}}(\sqrt{2}z^*)\bigg|^2.
 \label{eq:homogenous}
\end{align}
This equation is the crucial result of this subsection:  for parameters that bring us close to a quantum bistable point, it provides a simple way to understand the system's steady state and its extreme sensitivity to small parameter changes.

\begin{figure*}
     \centering
     \includegraphics[width=0.99\textwidth]{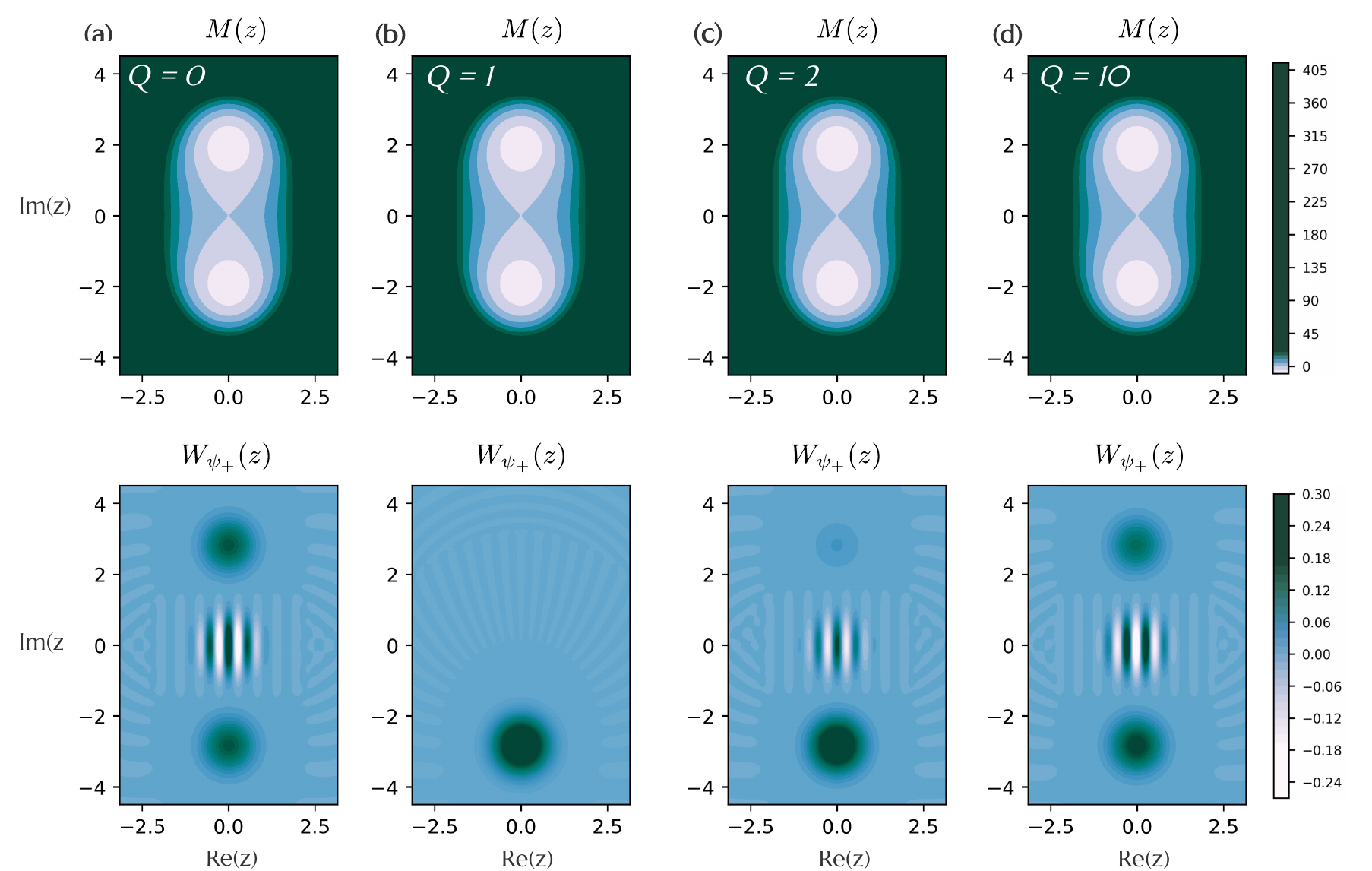}
     \caption{{\bf Extreme parameter sensitivity near a quantum-bistable point.}
     Bottom row: Wigner function for the purification $\psi_{+,\rm SB}$ of the Kerr-cavity steady state, for various parameter choices that are close to the $(r_1,r_2) = (0,0)$ quantum bistable point.  For all plots 
     $\Lambda_3 = \Delta = \kappa_2=0$, $\kappa_1=10^{-2}K$, and $\Lambda_2=4K$, and $\Lambda_1$ increases from left to right:    (a) $\Lambda_1=0$ ($Q=0$), (b), $\Lambda_1=0.01K$ ($Q=1$), (c) $\Lambda_1=0.02K$ ($Q=2$), and (d) $\Lambda_1=0.1K$ ($Q=10$).  The small value of $\kappa_1$ and $\Lambda_1$ imply that one is not exactly at the bistable point; the $Q$ parameter then controls the form of the unique steady state, c.f.~Eq.~(\ref{eq:Q}).  By tuning the single-photon drive amplitude $\Lambda_1$, one can pick out a particular superposition in the "bistable" manifold by varying $Q$. Top row:  corresponding metapotential $M(z)$ for the same parameter choices.  The metapotential is essentially unchanged for this range of $\Lambda_1$, showing that it cannot be used to understand the large changes in the quantum steady state.  
     }
     \label{fig:monotonic}
 \end{figure*}

\subsection{Cat-state bistability:  $(r_1,r_2) = (0,0)$}

The simplest bistable point is where $r_1 = r_2 = 0$.  From Eqs.~(\ref{eq:r1Definition}),(\ref{eq:r2Definition}), we see that this requires there to be no single photon drive or loss, nor any detuning:  $\Lambda_1 = \Delta = \kappa_1 = 0$. This corresponds to the well-known quantum bistability that occurs in a two-photon driven Kerr resonator \cite{Goto2016,puri_engineering_2017,Grimm2019}, a system where photon number parity is conserved.
  The two distinct solutions to the recurrence relation in Eq.~(\ref{eq:EffectiveRecurrence}) are $c_j = \delta_{j,0}$ and
$c_j = 1 - \delta_{j,0}$ (c.f.~Eqs.~(\ref{eq:bsphysics1})-(\ref{eq:bsphysics2})).  This corresponds to two distinct dark states for the $+$ mode, with SB wavefunctions
\begin{align}
    \psi_{1,\textrm{SB}}(z)&=
        e^{-\epsilon z}\label{eq:cat_states1}\\
    \psi_{2,\textrm{SB}}(z)&=
        e^{\epsilon z} - e^{-\epsilon z}
        \label{eq:cat_states2}
\end{align}
$\psi_{1,\textrm{SB}}(z)$ corresponds to a coherent state with amplitude $\epsilon  \equiv  i\sqrt{\lambda_2}$, whereas $\psi_{2,\textrm{SB}}(z)$ corresponds to an odd cat state (odd superposition of coherent states with amplitude $\epsilon$). Note that we have picked the global phase of $\psi_{2,\rm SB}$ to be compatible with 
Eq.~(\ref{eq:homogenous}).\\

We thus have a direct connection between this parity-based bistability and the photon blockade and anti-blockade discussed above:  bistability corresponds to both these phenomena occurring simultaneously.  As always, any amount of single-photon loss will kill the bistability and yield a unique steady state (though relaxation to this state could be extremely slow).  Our approach gives a simple way to understand the unique steady state when there is weak single photon loss, and possibly other weak perturbations (such as single photon driving and/or a detuning).  These imperfections cause a shift in the recursion parameters away from $\delta r_1 = \delta r_2 = 0$:
\begin{align}
    \delta r_2 & = 
        \frac{2\Delta + i \kappa_1}{K - i \kappa_2} \\
    \delta r_1 & = \frac{\delta r_2}{2} -
        \frac{i\Lambda_1}{\sqrt{ \Lambda_2 (K - i \kappa_2)}}
\end{align}
For small imperfections, we can then use Eq.~(\ref{eq:homogenous}) to give us the steady-state SB wavefunction:
\begin{equation}
    \psi_{+,\textrm{SB}}(z) =
    N
    \left[ (1+Q) e^{-\epsilon z} + (1-Q) e^{\epsilon z}) \right], \label{eq:DecoheredCatSB}
\end{equation}
where $N$ is a normalization constant, and
\begin{equation}
    Q = 
       \sqrt{\frac{K - i \kappa_2}{\Lambda_2/4}} 
        \frac{i\Lambda_1 }{2\Delta + i \kappa_1}.\label{eq:Q}
\end{equation}
Eq.~(\ref{eq:DecoheredCatSB}) directly gives us the Wigner function of the unique steady state via Eq.~(\ref{eq:SBWignert}).  Each term in Eq.~(\ref{eq:DecoheredCatSB}) on its own corresponds to a simple coherent state (amplitudes $ \pm \epsilon \equiv \pm i\sqrt{\lambda_2}$).  This equation also reveals something surprising: the localization of the steady state in phase space is a non-monotonic function of $\Lambda_1$.  The state is delocalized both for $\Lambda_1 = 0$, and for $\Lambda_1$ large enough to make $Q \gg 1$. Representative results are shown in Figure \ref{fig:monotonic}; 
We plot the semiclassical metapotential for in this figure for each parameter choice; it shows almost no changes, indicating that it cannot be used to understand the strong parameter-sensitivity of the quantum steady state.

\subsection{Quantum bistability with a single photon drive: $(r_1,r_2) = (n,n)$}

A more surprising regime of near bistability is when the recursion parameters are both tuned to be close to the same positive integer, 
i.e.~$(r_1,r_2) \simeq (n,n)$.  As discussed, for an exact tuning to this point, the expanded system exhibits quantum bistability.  There are two orthogonal solutions to the recurrence relations, given by
$c_j = \sum_{k=1}^n \delta_{j,k}$ and $c_j = \sum_{k=1}^n (1-\delta_{j,k})$
(c.f. Eq. (\ref{eq:bsphysics1}-\ref{eq:bsphysics2})). 
These in turn correspond to two distinct $+$-mode states\\
\begin{align}
    \psi_{1,\textrm{SB}}(z)&=N_1e^{-\epsilon z}\,\Gamma(n+1, 2\epsilon z)\label{eq:cat_states3}\\
    \psi_{2,\textrm{SB}}(z)&=N_2e^{-\epsilon z}\Bigg(1-\fr{\Gamma(n+1, 2\epsilon z)}{\Gamma(n+1)}\Bigg)\label{eq:cat_states4}
\end{align}
where $\Gamma(r,z)\equiv \int_z^\infty t^{r-1}e^{-t}dt$ is the incomplete Gamma function. 

In the absence of any loss, tuning $r_1 = r_2 = n$ requires a detuning $\Delta = n/2 K$ and a single photon drive $\Lambda_1 = -i(n/2) \sqrt{\Lambda_2 K}$.  If we now include single photon loss (but keep $\kappa_2 = 0$), and also shift $\Lambda_1$ slightly from the above value, the recurrence parameters are slightly shifted as well:
\begin{eqnarray}
    r_1 & = & n + 
    \frac{i \kappa_1}{2 K}
    \equiv n + \delta r_1 \\
    r_2 & = &
    n + \frac{i \kappa_1}{4 K}
    -i\frac{ \delta \Lambda_1 }{\sqrt{K \Lambda_2}}
    \equiv n + \delta r_2
\end{eqnarray}Hence, via Eq.~(\ref{eq:homogenous}), by slightly varying the one photon drive amplitude, one can pick out completely different linear combinations of the two different bistable states as the single unique steady state.  This leads to an extreme sensitivity of the final state to small changes in $\Lambda_1$. Note that by picking parameters so that $\delta r_1 = \delta r_2$, the steady state becomes a coherent state with amplitude $\gamma = \sqrt{- \Lambda_2/K}$, whereas if $\delta r_1 = 0$, it has a bimodal form.

\subsection{Metastability due to proximal quantum bistability}

\begin{figure}
      \centering
      \includegraphics[width=0.99\columnwidth]{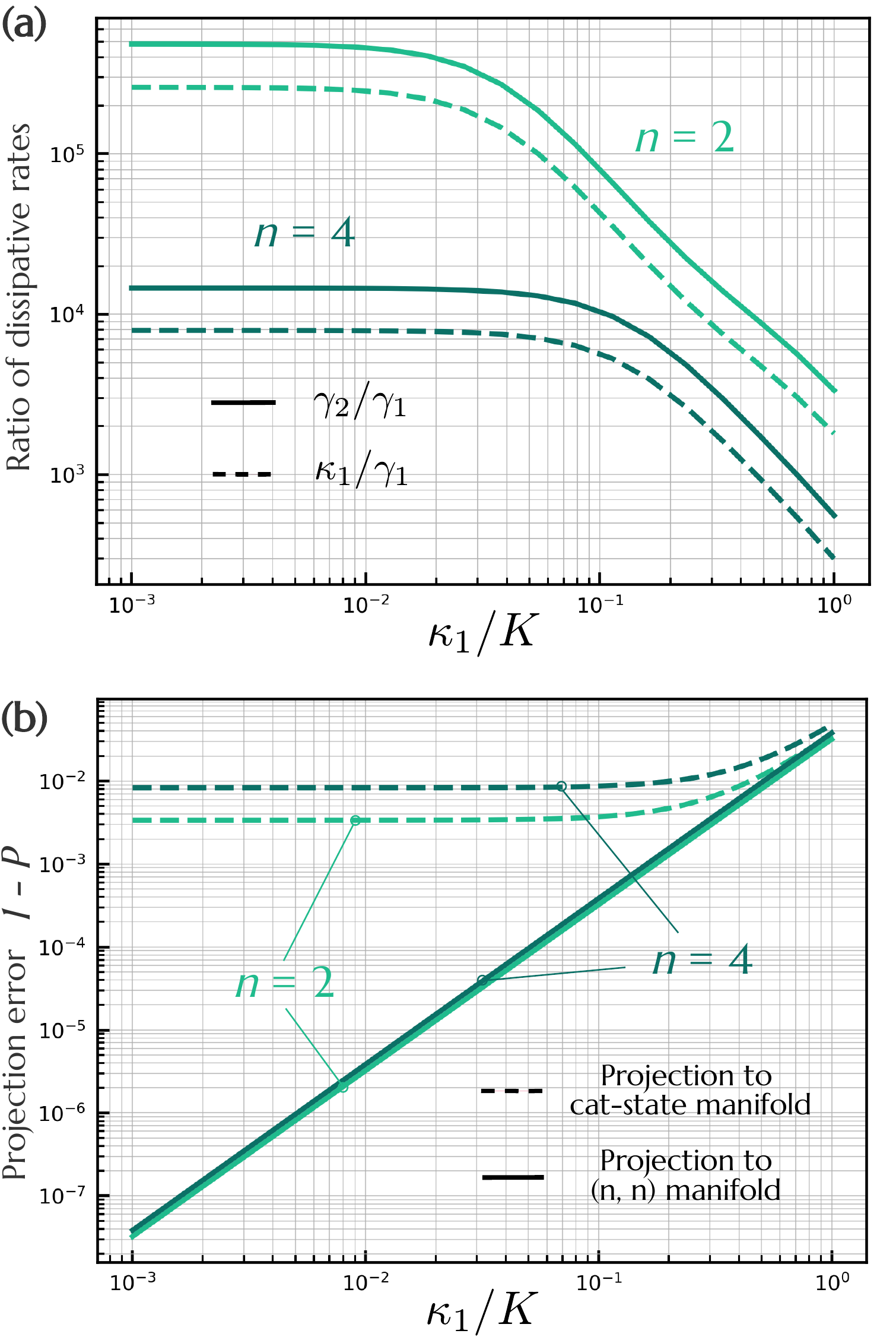}
      \caption{
      {\bf Slow dynamics near generalized bistable regimes.}
      (a)  Solid line: ratio of the two smallest relaxation rates (i.e.~dissipative rates of the system Liouvillian $\mathcal{L}_0$), as a function of $\kappa_1$.  Dashed line: $\kappa_1/\gamma_1$.  $\kappa_1 \rightarrow 0$ corresponds to being at a bistable parameter point $(r_1,r_2) = (n,n)$, 
      either $n=2$ (light green) or
      $n=4$ (dark green).
      Parameters are $\Lambda_2=6K$, $\Lambda_3 = \kappa_2=0$, $\Delta = nK/2$, and $\Lambda_1=-in\sqrt{\Lambda_2}/2$.
      One sees that the slow rate $\gamma_1$ is much slower than $\kappa_1$, and that there is a pronounced dissipative gap.  
      (b) Solid line: The measure $1-P$ 
      (c.f.~Eq.~(\ref{eq:error})) of how closely the slowest system decay mode (with rate $\gamma_1$) corresponds to dynamics in the bistable manifold. Dashed lines: same, but measuring how closely this mode is described by coherent states centered at the semiclassical stable amplitudes.  One clearly sees that the bistable manifold gives a far better description.  Same parameters as in (a).  }
      \label{fig:rates}
  \end{figure}

Tuning parameters to be close to a quantum bistable point also has consequences for dynamics. The characteristic decay rates of the system correspond to the non-zero eigenvalues of the Liouvillian $\mathcal{L}_0$ (c.f.~Eq.~(\ref{eq:one_mode})).  We find that tuning to a regime of near-bistability gives rise to an extremely slow population-decay mode, and also a clear dissipative gap separating the rate of this slow-mode from other decay modes. Formally, if we let $\gamma_j$ denote the decay modes of the Liouvillian (i.e.~negative real parts of the eigenvalues of $\mathcal{L}_0$), and order rates such that $\gamma_1 \leq \gamma_2 \leq ....$, then in near-bistable regimes:
\begin{equation}
    \gamma_1 \ll \kappa_1, \,\,\,\,\,
    \gamma_2 \gg \gamma_1\label{eq:separation}
\end{equation}
Note that this hierarchy of dissipative rates has already been described for the more familiar $(r_1,r_2) = (0,0)$ ``cat-state'' bistable point \cite{puri_engineering_2017}; we show that this is also true for our new bistable points.  
An exact description of this dynamical behaviour is outside the scope of the CQA method.  It can however be studied numerically.  
Representative behavior of a driven Kerr cavity whose parameters are close to either the $(r_1,r_2) = (2,2)$ or $(4,4)$ bistable points are shown in 
Fig.~\ref{fig:rates}(a).

For near-bistable parameters, the CQA approach provides insight into the nature of the slow decay mode of $\mathcal{L}_0$. As one might expect, this mode corresponds to slow relaxation within the bistable manifold of states. For more general works on metastability in open quantum systems, see \cite{macieszczak_towards_2016,kessler_dissipative_2012}. To make this precise, recall that if one tuned exactly to a bistable parameter point, cavity-$a$ has a continuous three-parameter family of possible steady states corresponding to  Eq.~(\ref{eq:BistableSteadyState}) (and incoherent mixtures of these states).  Density matrices in this bistable manifold lie in the span of the four operators ($i,j = 1,2$):
\begin{align}
    \hat{M}_{ij} &=
    \text{tr}_b \left[
        \left( 
            \ket{\psi_i}\bra{\psi_j} \right)_+
        \left( 
            \ket{0}\bra{0} \right)_-
    \right]
        \label{eq:Mij}
\end{align}
By Appendix \ref{app:wigner}, these operators have Wigner transforms
\begin{align}
W_{ij}(z)&=N\psi_{i,\text{SB}}(\sqrt{2}z^*)\psi_{j,\text{SB}}^*(\sqrt{2}z^*)e^{-2|z|^2}\label{eq:modes}
\end{align}
with $N$ a normalization constant.

The slow mode (rate $\gamma_1$) has an associated right eigenvector $\hat{M}_{\rm slow}$, i.e. $\mathcal{L}_0 \hat{M}_{\rm slow} = - \gamma_1 \hat{M}_{\rm slow}$.  If the slow dynamics is entirely in the bistable manifold, then $\hat{M}_{\rm slow}$ should lie completely within the span of the $\hat{M}_{ij}$.  
To see whether this is the case, we pick parameters for near-bistability, and numerically calculate the Hilbert-Schmidt norm $P$ of the projection of the slow mode onto the bistable subspace:  
\begin{align}
P \equiv \sum_{i,j=1,2}|\text{Tr}[\widetilde{M}_{ij}^{\dag}\hat{M}_\text{slow}]|^2\label{eq:error}
\end{align}
Here $\widetilde{M}_{ij}$ is an orthonormal basis for the span of $\hat{M}_{ij}$ (obtained via the standard Gram-Schmidt process).  As $0 \leq P \leq 1$, the quantity $1 - P$ measures how much of the slow mode's dynamics lies outside the bistable manifold.

Representative results for $1 - P$  are shown in Fig.~\ref{fig:rates}(b).  One sees that for small $\kappa_1$ (i.e.~when one is close to the bistable point), the slow mode is almost entirely described by the bistable state manifold. For comparison, we have also tried to describe the slow mode in terms of simple coherent states centered at the expected classical bistable steady-state amplitudes.  This involves taking
\begin{align}
    \hat{M}^{ij}_\text{cat}&=|\alpha_i\rangle \langle \alpha_j|
\end{align}
with $\alpha_j$ the classical amplitudes, determined by  
(with $K\equiv 1$):
\begin{align}
\Delta(\alpha_j-i\sqrt{\Lambda_2})-\Lambda_2\alpha^*_j-\alpha_j|\alpha_j|^2\equiv 0,~~~j=1,2
\end{align}
One sees from Fig.~\ref{fig:rates}(b) that this coherent-state description does a far poorer job of describing the dynamical slow mode, compared to the states from the bistable manifold. While metastability in two-photon driven Kerr resonators was discussed in \cite{minganti_spectral_2018},  its connection to the existence of nearby, novel quantum bistable points  (i.e. generalized cat-state regimes) in the resonator's phase diagram has not been previously investigated.

\section{Parity-conserving dynamics: true quantum bistability}
\label{sec:bistability}

We now focus on a special case that has received considerable recent attention \cite{mirrahimi_dynamically_2014,Goto2016,puri_engineering_2017,Grimm2019}: a system where $\kappa_1 = \Lambda_1 = 0$ in Eq.~(\ref{eq:one_mode}), implying that the full dynamics conserves photon number parity.  This in turn implies that there are at least two distinct steady states, and opens the possibility of true quantum bistability. Note that a comprehensive discussion of generic Lindblad master equations with multiple steady states is provided in Ref.~\cite{Albert2014}. 
Our exact-solution CQA method provides several insights into this regime. Among other things, it allows one to understand why adding a drive-detuning destroys quantum bistability despite parity still being conserved, something that is not possible with $P$-function methods, which give a unique solution \cite{hach_iii_generation_1994, gilles_generation_1994}. In addition, the CQA method also gives a succinct analytical expression that controls which unique steady state is selected from the bistable manifold when quantum bistability is broken.

We start by revisiting the CQA method of Sec.~\ref{sec:the absorber method} for systems described by Eq.~(\ref{eq:one_mode}) with $\kappa_1 = \Lambda_1 = 0$. 
The corresponding cascaded two-cavity system is described by Eqs.~(\ref{eq:CascadedMEQ_2}) and (\ref{eq:cascaded_ham2}).  
The first step as always is to insist that we have a state that is dark with respect to the cascaded dissipators.  For $\kappa_1 = 0$, we only have the a two-photon loss dissipator, given by
\begin{equation}
\mathcal{D}[\hat{a}^2 - \hat{b}^2] = \mathcal{D}[2 \hat{c}_+ \hat{c}_-]
\end{equation}
where again the collective $\hat{c}_{\pm}$ modes are defined in Eq.~(\ref{eq:cmodes}).  There are now two distinct possibilities for a non-trivial dark state:
either the $\hat{c}_-$ mode is forced to be in vacuum (with the $+$ mode occupied), or the $\hat{c}_+$ mode is forced to be in vacuum (with the $-$ mode occupied).  The first option is the same as what we did for $\kappa_1 = 0$; the second option is a new possibility enabled by the lack of one photon loss.

It follows that the most general 2-cavity dark state has the form:
\begin{equation}
    \ket{\psi_{\rm dk}} = \alpha_+ \ket{\psi}_+ \ket{0}_- + \alpha_- \ket{0}_+ \ket{\theta}_-
    \label{eq:ParityDark}
\end{equation}
This structure is a direct consequence of parity conservation, which guarantees the existence of at least two orthogonal steady states (one even parity, one odd parity).  This structure also implies that the general argument in Sec.~\ref{sec:the absorber method} ensuring a positive cavity-$a$ steady-state Wigner function no longer holds, as $\ket{\psi_{\rm dk}}$ can have both $+$ and $-$ modes occupied.

\subsection{Zero detuning: quantum bistability}

Consider first the case $\Delta = 0$, meaning that we have a resonantly-driven Kerr parametric oscillator subject to two photon loss.  Without dissipation, this system has degenerate coherent state eigenstates \cite{Goto2016,puri_engineering_2017}.  Including two-photon loss, the dissipative system exhibits true quantum bistability: the steady-state manifold corresponds to a two-dimensional decoherence-free subspace, in the language of \cite{albert_lindbladians_2017}. We show how this structure emerges via the CQA method.

The first step of the CQA method is to identify possible pure dark states of the collective cascaded-systems dissipators; in our case, this is Eq.~(\ref{eq:ParityDark}).  To have this state be a steady state, it must also be an eigenstate of the cascaded Hamiltonian (c.f.~Eq.~(\ref{eq:cascaded_ham2})).  For $\Delta = 0$, this leads to the equation:
\begin{align}
    \hat{c}_-^\dag \hat{c}_+^\dag \Bigg(
        \alpha_+ \left[ \hat{c}_+^2+\lambda_2 \right] \ket{\psi}_+ \ket{0}_{-} &&
        \nonumber \\
        + \, 
        \alpha_{-} \left[\hat{c}_-^2+\lambda_2 \right] \ket{0}_+ \ket{\theta}_{-} 
        \Bigg)
        = E \ket{\psi_{\rm dk}} &&
\end{align}
with $\lambda_2 = 2 \Lambda_2 / (K - i \kappa_2)$.
Since $\hat{H}_\text{casc}$ always adds an excitation to {\it both}  modes $\hat{c}_-$ and $\hat{c}_+$, the only possible energy eigenvalue is $E=0$.  The equation then decouples into separate equations for $\ket{\psi}_+,\ket{\theta}_-$ which are easily solved.  Crucially, each of these equations admits two possible solutions.

As a result, one finds that the most general dark state solution can be written in terms of coherent states as:
\begin{align}
    \ket{\psi_{\rm dk}} & = 
        \sum_{\pm} \left(
            \mu_{\pm} \ket{\pm  \epsilon}_+ \ket{0}_- + 
            \nu_{\pm} \ket{0}_+ \ket{\pm  \epsilon}_- 
            \right) 
            \label{eq:ParityDarkState}
\end{align}
where the coherent state amplitude $\epsilon = i\sqrt{\lambda_2}$, as in the previous section.  We see that the cascaded two-cavity system has a four dimensional subspace of possible steady-state, dark states.

The last step is to determine the corresponding steady state structure of the physical $a$ cavity.  As discussed in Sec.~\ref{sec:the absorber method}, this effectively corresponds to taking a given two-cavity state, sending it through a 50-50 beamsplitter, and then discarding one of the outputs.  This procedure is easy to carry out on the general state in Eq.~(\ref{eq:ParityDarkState}), as coherent states transform in a simple manner under a beamsplitter operation.  In the basis of the physical $a$ cavity and auxiliary $b$ cavity, our general dark state has the form:
\begin{align}
    \ket{\psi_{\rm dk}} & = 
        \sum_{\pm} \Big(
            \mu_{\pm} \ket{\pm  \tilde{\epsilon}}_a \ket{\pm  \tilde{\epsilon}}_b \nonumber\\
&~~~~~~~~~~~~~~~~+ 
            \nu_{\pm} \ket{\pm  \tilde{\epsilon}}_a \ket{\mp  \tilde{\epsilon}}_b \Big) \\
    & = 
        \Big(
            \mu_{+} \ket{  \tilde{\epsilon}}_a + 
            \nu_{-} \ket{  \tilde{\epsilon}}_a 
        \Big) \ket{  \tilde{\epsilon}}_b \nonumber\\
        &~~~~~+
        \Big(
            \mu_{-} \ket{ - \tilde{\epsilon}}_a + 
            \nu_{+} \ket{ + \tilde{\epsilon}}_a 
        \Big) \ket{ - \tilde{\epsilon}}_b,
\end{align}
with $\tilde{\epsilon} \equiv \epsilon/\sqrt{2}$. As there is in general entanglement between the physical $a$ cavity and the auxiliary $b$ cavity, one in general is left with an impure state for cavity $a$.  However, pure cavity-$a$ steady states are indeed possible; consider for example the case where $\mu_- = \nu_+ = 0$.

The upshot is that we have a steady state manifold for cavity $a$ that is two dimensional, and spanned by the states $\ket{\pm \tilde{\epsilon}}_a$ (in agreement with previous work \cite{mirrahimi_dynamically_2014}).  In simple terms, the steady-state manifold corresponds to a quantum bit, i.e.~a full single-qubit Bloch sphere \cite{Albert2014}.  This is what we mean by the system exhibiting quantum bistability.

\subsection{Non-zero detuning: classical bistability}

\subsubsection{Loss of quantum bistability}
We next consider the case of a adding a non-zero detuning $\Delta$.  As has been discussed previously \cite{mirrahimi_dynamically_2014}, this causes the steady-state manifold to transition from being a two-dimensional decoherence free subspace (i.e.~quantum bistability) to having the structure of a simple classical bit.  Formally, it corrsponds to an orthogonal direct sum of two one-dimensional noiseless subsystems (one for each parity sector). We will find that, in contrast to $P$-function methods \cite{minganti_exact_2016}, CQA is able to analytically detect this transition, and gives closed-form expressions for each of the direct-summands in the steady-state manifold. Solving the system again using the CQA method, the requirement of having our general dark state in Eq.~(\ref{eq:ParityDark}) be a energy eigenstate of the cascaded Hamiltonian leads to the equations
\begin{align}
    \hat{c}_-^\dag \left( 
        \hat{c}_+^\dag\hat{c}_+^2 - D\hat{c}_+ +\lambda_2 \hat{c}_+^\dag \right)
        |\psi\rangle_+&=0,\\
\hat{c}_+^\dag\left( \hat{c}_-^\dag\hat{c}_-^2-D\hat{c}_- + \lambda_2 \hat{c}_-^\dag\right)|\theta\rangle_-&=0.
\end{align}
where $D = 2\Delta / (K - i \kappa_2)$. As before, the equations determining $\ket{\psi}_+$ and $\ket{\theta}_-$ are identical (reflecting parity conservation).  The equation in each case can be solved by using a SB representation for the state, and turning the operator equations into differential equations. We get the same ODE in each case:
\begin{align}
    \Big(z\pdd{}{z}-D\pd{}{z}+\lambda_2z\Big)\psi_{\rm SB}(z)=0
\end{align}
With the same equation for $\theta_{\rm SB}$, and with $\lambda_2,D$ having the same definitions as earlier in the main text. 

At the qualitative level, one can see how true quantum bistability is lost in the presence of detuning: for zero detuning $D\equiv 0$, the ODE above has no singular points, and thus the standard existence theorem (\S 12.22 in \cite{ince_ordinary_1956}) guarantees two independent, analytic solutions. As discussed earlier, this leads to quantum bistability for the physical mode $a$. However, the term $\propto D\partial_z$ introduces a singular point into the ODE at $z=0$, and the existence of two dark steady states is no longer guaranteed. Indeed, the singular point at $z=0$ produces a branch-cut discontinuity in one of the solutions. Generically, only one analytic solution survives:
\begin{align}
\psi_\text{SB}(z)&=\frac{1}{N^{1/2}}\,_0F_1(1/2-D/2;-\lambda_2z^2/4),\label{eq:nonsingular}
\end{align}
where  $N$ is a normalization constant. 

As we will see, this two-fold reduction in the number of dark steady-states has dramatic consequences for the bistability of the physical mode $a$. As there is a unique choice for both $\ket{\psi}_+$ and $\ket{\theta}_-$, the most general dark state has the form of Eq.~(\ref{eq:ParityDark}), and corresponds to a two-dimensional subspace.  In what follows, it will be useful to write this general dark state as 
\begin{equation}
    \ket{\psi_{\rm dk}} = \mu_{\rm e} \ket{\Phi_{\rm e}} + \mu_{\rm o} \ket{\Phi_{\rm o}}
    \label{eq:DetunedDarkState}
\end{equation}
with
\begin{align}
      \ket{\Phi_{\rm e/o}} & = 
        \frac{1}{\sqrt{2 \pm 2 N^{-1}}}
        \left( \ket{\psi}_+ \ket{0}_- \pm \ket{0}_+ \ket{\psi}_- \right). \label{eq:eo}
\end{align}

We now trace-out the auxiliary $b$ cavity.  Note that the pure dark states above span a subspace of dimension 2.  Incoherent mixtures in this subspace are also stationary states; hence the cascaded 2 cavity problem has a steady-state manifold corresponding to a Bloch sphere.  We imagine starting with an abitrary mixed state in this subspace (described by a 2 cavity density matrix), and then tracing out cavity $a$ to determine the corresponding cavity $a$ state.  Understanding the full range of cavity $a$ states produced here determines the steady-state manifold of cavity $a$.

This procedure leads us to consider four linearly-independent cavity-$a$ operators (that determine the cavity $a$ density matrix after tracing out cavity $b$):
\begin{align}
    \hat{M}^{++}_{a,ss} &\equiv \text{tr}_b[|\psi\rangle_+ |0\rangle_- \langle \psi|_+ \langle 0|_-]\nonumber \\
    \hat{M}^{--}_{a,ss} &\equiv \text{tr}_b[|0\rangle_+ |\psi\rangle_- \langle 0|_+ \langle \psi|_-]\nonumber\\
    \hat{M}^{+-}_{a,ss} &\equiv \text{tr}_b[|\psi\rangle_+ |0\rangle_- \langle 0|_+ \langle \psi|_-]\nonumber\\
    \hat{M}^{-+}_{a,ss} &\equiv (\hat{M}^{+-}_{a,ss})^\dag\label{eq:rho_plus_origin}
\end{align}
To understand the structure of these operators, we consider their corresponding $Q$-functions (easily obtainable using the SB representation):
\begin{align}
    Q^{\pm\pm}_{a,ss}(z)&=\int d^2u \,\psi_{\rm SB}^*\Big(\fr{z\pm u}{\sqrt{2}}\Big)\psi_{\rm SB}\Big(\fr{z\pm u}{\sqrt{2}}\Big)\\ 
    Q^{\pm\mp}_{a,ss}(z)&=\int d^2u \,\psi_{\rm SB}^*\Big(\fr{z\pm u}{\sqrt{2}}\Big)\psi_{\rm SB}\Big(\fr{z\mp u}{\sqrt{2}}\Big) 
\end{align}
We obtain an important result: these four operators are not all independent.  
Because of the symmetry of each integral under the mapping $u\to -u$, we have 
\begin{align}
    \hat{M}^{++}_{a,ss}=\hat{M}^{--}_{a,ss},~~~~~\hat{M}^{+-}_{a,ss}=\hat{M}^{-+}_{a,ss}=(\hat{M}^{+-}_{a,ss})^\dag
\end{align}

These equalities imply a loss of information in tracing out cavity-$b$, and result in the cavity-$a$ steady state manifold being simply two-dimensional.  It is spanned by the quantities
\begin{align}
    \hat{\rho}_{a,ss}^+&\equiv \hat{M}_{a,ss}^{++}\nonumber\\ 
    \hat{\rho}_{a,ss}^-&\equiv \hat{M}_{a,ss}^{+-}.\label{eq:rho_plus}
\end{align}
with $\hat{\rho}_{a,ss}^\pm$ both Hermitian. We now have enough information to calculate each steady state exactly: since the steady-state manifold is two-dimensional, and since parity is a conserved quantity, every density matrix in the manifold must then be an impure mixture of the form
\begin{align}
    \hat{\rho}_{a,ss}&=p\hat{\rho}_e+(1-p)\hat{\rho}_o\label{eq:type_of_bistability}
\end{align}
where the extremal states $\hat{\rho}_{e/o}$ are uniquely characterized by the property of having definite photon number parity (even and odd respectively). 

Thus, in summary, in this case there is a distinct steady state in both the even and odd photon number sectors; any mixture of these is also a possible steady state.  The steady state manifold is indexed by just a single number $0 \leq p \leq 1$, which simply corresponds to the dynamically-conserved probability of having an even photon number parity.  In simpler terms, the cavity-$a$ steady-state manifold corresponds to a classical bit \cite{Albert2014}. 
To conclude our discussion of $\Delta \neq 0$, we use the CQA method to compute exactly each steady state in the bistable manifold. We begin by noting that the states $|\Phi_{e/o}\rangle$ in Eq. (\ref{eq:eo}) have definite photon-number parity, and thus so do the corresponding states of the physical cavity-$a$ (obtained by tracing over cavity-$b$). Therefore, by uniqueness of the extremal states, these states must be precisely $\hat{\rho}_{e/o}$:
\begin{align}
    \hat{\rho}_e& =\text{tr}_b[|\Phi_e\rangle\langle\Phi_e|]\nonumber\\
    \hat{\rho}_o& =\text{tr}_b[|\Phi_o\rangle\langle\Phi_o|].\label{eq:extremal}
\end{align}
To compute these steady-states, we note that by substituting Eq. (\ref{eq:eo}) into Eq. (\ref{eq:extremal}) we can expand, e.g. 
\begin{align}
\hat{\rho}_{e/o}&=\fr{N}{N\pm 1} (\hat{\rho}_{a,ss}^+\pm \hat{\rho}_{a,ss}^-)
\end{align}
where $N$ is just the normalization constant $N$ for the dark state $|\psi\rangle_+$, which has the exact expression
\begin{align}
N&=\,_1F_2(1/2;1/2-D/2,(1/2-D/2)^*;|\lambda_2/2|^2).\label{eq:normalization2}
\end{align}
Inverting the above linear relation, we get
\begin{align}
\hat{\rho}^+_{a,ss}&=\fr{1}{2}\left[\fr{N+1}{N}\hat{\rho}_e+\fr{N-1}{N}\hat{\rho}_o\right]\label{eq:bistable_manifold} 
\end{align}
This equation immediately leads to exact expressions for $\hat{\rho}_{e/o}$, which are given in Appendix \ref{app:TracingOut}. Furthermore, by comparison with Eq. \eqref{eq:rho_plus_origin}, $\hat{\rho}^+_{a,ss}$ also happens to be the steady-state in the presence of an infinitessimal amount of bistability-breaking single-photon loss. Therefore, in Eqs.  (\ref{eq:normalization2}-\ref{eq:bistable_manifold}), CQA is able to smoothly describe the transition from a Kerr oscillator having two quantum steady states to having only one. In the weak-driving limit $\lambda_2\to 0$, the hypergeometric series defining $N$ collapses to just the first term (c.f. Eq. \eqref{eq:normalization2}), and we get $N\to 1$, so
\begin{align}
\hat{\rho}_{a,ss}^+\underset{\lambda_2\to 0}{\sim}\hat{\rho}_e.
\nonumber
\end{align}
In contrast, in the strong-driving limit $N$ diverges, and thus
\begin{align}
\hat{\rho}_{a,ss}^+\underset{\lambda_2\to \infty}{\sim}\fr{\hat{\rho}_e+\hat{\rho}_o}{2}.
\end{align}
A final piece of physical intuition: since $N$ is a function only of the modulus $|\lambda_2|$, the relative bias (towards either $\hat{\rho}_{o/e}$) is independent of the phase $\phi$ of the drive $\lambda_2\equiv e^{i\phi}|\lambda_2|$.

\section{Conclusions}
\label{sec:conclusions}

In this, work, we have presented a generalization of the coherent quantum absorber method developed by Stannigel et. al. \cite{stannigel_driven-dissipative_2012} for solving the simplest driven Kerr resonator problem.  Our generalization exploited the Segal-Bargmann representation, and allows one to analytically solve for the steady state of driven-dissipative Kerr cavity models with nonlinear driving and nonlinear loss.  We used these analytic solutions to describe a host of new physical phenomena, including generalized photon-blockade phenomena, and new regimes of near quantum bistability.  These phenomena should be experimentally accessible in a number of different platforms, including superconducting circuit experiments.

Our work naturally suggests many new open questions and directions for future study.  For example, can the new bistable parameter points we have identified be utilized for quantum-information applications?  Are there other forms of nonlinear dissipation and driving that could also be included in our system that still leave it amenable to solution via the CQA method?  Can this approach be extended to nonlinear driven-dissipative systems with more than one cavity?  

At a fundamental level, there is also the basic question of {\it why} the CQA method is able to yield exact solutions to systems that are on the surface highly non-trivial (because of strong nonlinearities and driving).  Is there some general physical principle here, or perhaps a dissipative version of integrability that underlies this method?  These are all questions we hope to explore in future works.



\begin{acknowledgments} This work was supported by the Air Force Office of Scientific Research MURI program, under grant number FA9550-19-1-0399. 
\end{acknowledgments}


\appendix

\section{Circuit QED realization of the model}
\label{app:cQED}

We now show that it is possible to realize the generalized driven-Kerr oscillator using cQED devices that already currently exist, in particular, a superconducting nonlinear asymmetric
inductive element (a.k.a. SNAIL device) \cite{sivak_kerr_2019}.  The Hamiltonian for a SNAIL, as a function of applied magnetic flux, can be written as (following the notation of \cite{sivak_kerr_2019}):
\begin{align}
    \hat{H}=\omega \hat{a}^\dag \hat{a}+ g_3(\Phi)(\hat{a}+\hat{a}^\dag)^3+g_4(\Phi)(\hat{a}+\hat{a}^\dag)^4.
\end{align}
We now introduce time-dependence into the flux parameter $\Phi$ in such a way that $g_3(t)\equiv g_3(\Phi(t))$ is oscillating at the cavity frequency with amplitude $g_3^{(0)}$, whereas $g_4(t) \equiv g_4^{(0)}$ is essentially constant (c.f. Fig. 1 in \cite{sivak_kerr_2019}). In the frame rotating at the cavity frequency, the time-dependent Hamiltonian then has the form
\begin{align}
    \hat{U}\hat{H}\hat{U}^\dag&= 3g_3^{(0)}\{ (\hat{a}^\dag \hat{a}^\dag \hat{a}+ \hat{a}^\dag \hat{a}\hat{a})+ (\hat{a}^\dag+\hat{a}) \}\nonumber\\
    &+3g_4^{(0)}\{2\hat{a}^\dag\hat{a}^\dag \hat{a}\hat{a}+ 4 \hat{a}^\dag \hat{a} + 1 \}\nonumber\\
    &~~~~~~~~~~~~~~~~~~~~~~+ \text{counter-rotating terms.}    
\end{align}
Under the rotating-wave approximation, if the drives are weak we can neglect all counter-rotating terms, which yields the effective Kerr Hamiltonian 
\begin{align}
    \hat{H}_{\rm RWA} &=\frac{K}{2}\hat{a}^\dag\hat{a}^\dag\hat{a}\hat{a}- \Delta \hat{a}^\dag \hat{a}\nonumber\\
    &+ (\Lambda_1\hat{a}^\dag +\Lambda_3\hat{a}^\dag\hat{a}^\dag \hat{a}+h.c.),
\end{align}
where $\Lambda_1 = \Lambda_3 = 3g_3^{(0)}$, and $K=12g_4^{(0)}=-\Delta$. In conclusion, realization of the nonlinear coherent driving effect, for weak driving strengths, is possible using a superconducting nonlinear asymmetric
inductive element, by modulating its flux parameter at the cavity frequency. We can also see from this analysis how it would be even harder to realize the three-photon additional / removal terms $(\hat{a}^\dag)^3,\hat{a}^3$ within this scheme, as this would require modulating the external flux $\Phi$ {\it three-times} more rapidly (specifically: 18 GHz, for the device considered in \cite{sivak_kerr_2019}).

\section{Exact realization of new quantum bistable regimes using a two-cavity non-cascaded setup}
\label{app:ExactBistability}

In Sec.~\ref{subsec:Bistability}, we discussed how the generalized driven-dissipative Kerr problem could be tuned to be arbitrarily close to points in parameter space where we have true quantum bistability.  Exact tuning to a bistable point was not possible due to the constraint that neither $\kappa_1$ nor $\kappa_2$ could be made negative.

\begin{figure}[t]
      \centering
      \includegraphics[width=0.99\columnwidth]{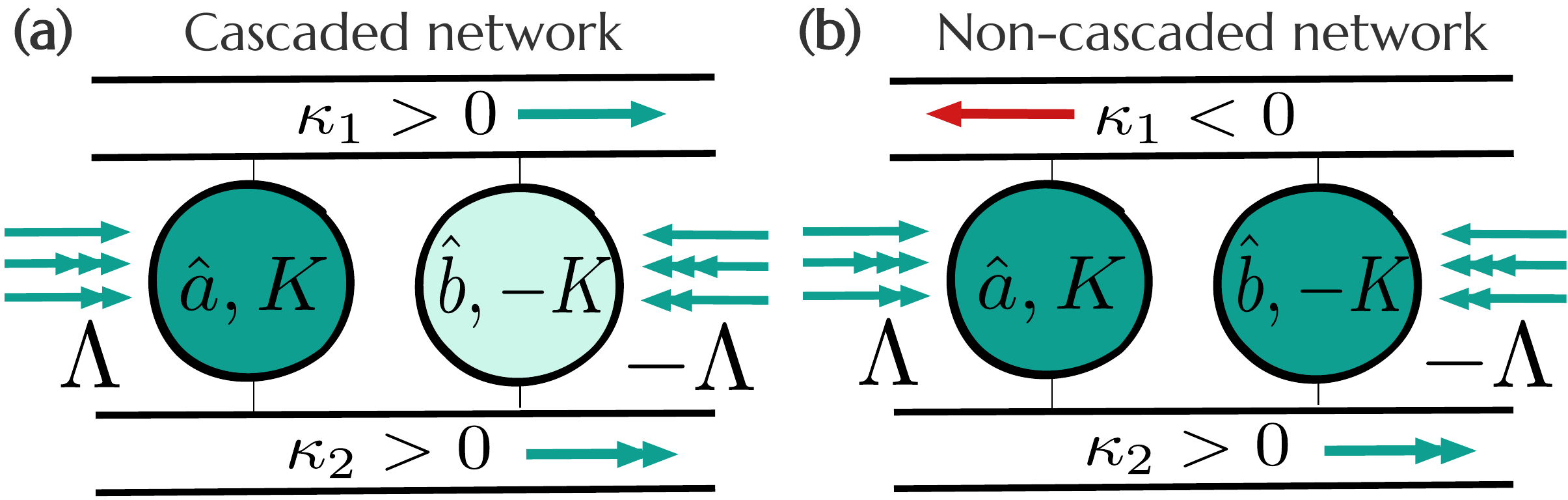}
      \caption{\textbf{Realizing quantum bistability by breaking chirality}. The generalized bistable points in our phase diagram (c.f.~Fig.~\ref{fig:phase_diagram}) are {\it exactly} realizable by using a two-cavity setup (b) which is not cascaded, i.e. where the chirality of one of the waveguides is reversed.}
      \label{fig:noncascaded}
  \end{figure}

An exact realization of these quantum bistable points is nonetheless possible if one works with the two cavity system in Fig.~\ref{fig:virtual_cavity_intro}.  Making one of $\kappa_1$ or $\kappa_2$ negative now has a simple physical interpretation:  we simply {\it reverse} the chirality of one of the waveguides in the absorber setup (see Fig.~\ref{fig:noncascaded}).  
Reversing the chirality of the (e.g. linearly-coupled) waveguide leads to the dynamics of the master equation Eq. (\ref{eq:two_mode}) with the same dissipators but with the Hamiltonian (c.f. Eq. (\ref{eq:cascaded_ham})) changed to
\begin{align}
    \hat{H}_{ab} 
        \to \hat{H}_a - \hat{H}_b +\fr{i\kappa_1}{2}(\hat{a}^\dag \hat{b}-h.c.)-\fr{i\kappa_2}{2}(\hat{a}^\dag \hat{a}^\dag \hat{b}^2-h.c.).
        \label{eq:cascaded_ham_revised}
\end{align}
Again, using the absorber method, we can solve this master equation in a manner identical to before, i.e. with $|\psi\rangle =|\psi_+\rangle|0_-\rangle$,  except now we have $\kappa_1\to -\kappa_1$. So the master equation specified by Eq. (\ref{eq:cascaded_ham_revised}) constitutes an analytic extension of the steady state to negative values of $\kappa_{1,2}$ (and thus arbitrary values of $D$), and thus can exhibit quantum bistability. For a depiction of the setup, see Fig. (\ref{fig:noncascaded}). 

In this case Eq. (\ref{eq:bistable_SB}) actually becomes a relation for selecting a pure state in the bistable manifold:
\begin{align}
|\psi_+\rangle&=\delta r_2|\psi_{+,1}\rangle +\delta r_1|\psi_{+,2}\rangle.
\end{align}
In this case, $|\psi_{+,j}\rangle$ are the photon-added coherent states of the symmetric mode defined in Sec. \ref{sec:the absorber method}. The states are perhaps best understood in the Fock basis. For $\Lambda_3=0$, 
\begin{align}
|\psi_{+,1}\rangle &= \sum_{~m=0~~}^{n}\fr{(2\epsilon\hat{c}_+^\dag)^m}{m!}|-\epsilon \rangle\\
|\psi_{+,2}\rangle &= \sum_{m=n+1}^{\infty}\fr{(2\epsilon\hat{c}_+^\dag)^m}{m!}|-\epsilon \rangle
\end{align}
where $|z\rangle$ as usual denotes a coherent state with amplitude $z$. Note that their sum {\it is} Gaussian, i.e. a coherent state, as is expected from properties of Kummer's hypergeometric function. In the more general case of the ''off-diagonal" bistable points (i.e. the $(n,m)$ points with $n\neq m$), one stabilizes even more exotic states, whose sum may no longer be Gaussian.

\section{Steady-state Wigner function}
\label{app:wigner}

 We will now rigorously prove the connection between the Wigner function of the steady-state $\hat{\rho}_{a,ss}$ of the driven Kerr cavity, and the modulus squared of the SB representation of its purification $|\psi_{+}\rangle$. We will also show how the calculation generalizes to the case where there are multiple dark states. The result here relies on a deep fact relating operator ordering conventions for a quantum-mechanical mode, and the heat semigroup on the corresponding classical phase space. This was originally pointed out by Glauber and Cahill in \cite{cahill_density_1969}, and we will review the salient results here. Specifically, given a (possibly non-Hermitian operator) $\hat{A}$ of a quantum-mechanical mode, define its {\it normally-ordered symbol} $\sigma_N$ to be
 \begin{align}
      \sigma^N_A(z) := :\hat{A}:\big|_{\hat{a}^\dag,\hat{a}\mapsto z,z^*},
 \end{align}
 where $:\hat{A}:$ is the operator $\hat{A}$ but re-expressed in normal order, i.e. with all of the creation operators to the left of the annihilation operators. Analogously, we can define the {\it symmetrically-ordered symbol}:
 \begin{align}
      \sigma_A(z) := :\hat{A}:_S\big|_{\hat{a}^\dag,\hat{a}\mapsto z,z^*},
 \end{align}
 where $:\hat{A}:_S$ is the operator $\hat{A}$ but re-expressed according to the symmetric ordering convention (as defined in \cite{cahill_density_1969}). The symmetrically-ordered symbol is proportional to the standard {\it Wigner transform}, which can be formally computed via an integral:
 \begin{align}
     \sigma_A(z)&\propto \int d^2\xi\, \text{Tr}[e^{\xi^* \hat{a} -\xi \hat{a}^\dag}\hat{A}]e^{\xi z^*-\xi^* z}
 \end{align}
 For positive semi-definite operators, e.g.  a density matrix $\hat{\rho}\equiv \hat{\rho}^\dag$, the symmetrically-ordered and normally-ordered symbols coincide with the Wigner- and Q-functions respectively:
 \begin{align}
     Q(z)=\frac{1}{\pi}\sigma_\rho^N(z),~~~~~~~W(z)=\frac{1}{\pi}\sigma_\rho(z).
 \end{align}
 What Glauber and Cahill showed in \cite{cahill_density_1969} is that operator symbols corresponding to different ordering conventions are related by the heat semigroup. In particular, we have the following theorem:
 
\noindent {\bf Theorem} (\cite{cahill_density_1969}). {\it Let $\hat{A}$ be a Hilbert-Schmidt operator (i.e. $\text{Tr}[\hat{A}^\dag \hat{A}]<\infty$). Then its normally-ordered symbol can be obtained by ''cooling" (i.e. running the heat equation on) the symmetrically-ordered symbol for a time $t=1/8$, i.e.}
 \begin{align}
     \sigma_A^N(z) = \int_{\mathbb{C}}d^2z' \mathcal K(t,z,z') \sigma_A(z') \bigg|_{t=1/8}
 \end{align}
 {\it where $K(t,z,z')$ is the heat kernel on $\mathbb{C}$, which can be exactly computed and comes out to }
 \begin{align}
     \mathcal K(t,z,z') = \frac{e^\frac{-|z-z'|^2}{4t}}{4\pi t}.
 \end{align}
 We can use the theorem above to directly trace-out the ancilla cavity used in the absorber method in the main text. Following the notation of the main text, suppose we have an orthonormal basis $|\psi_1\rangle, \cdots, |\psi_{k}\rangle$ of the space of dark states, i.e.
\begin{align}
    \hat{c}_-|\psi_{j}\rangle =0.
\label{eq:pure_states}
\end{align}
Clearly, these states form a $k$-dimensional Bloch sphere, spanned by their outer-products:
\begin{align}
    |\psi_{i}\rangle\langle \psi_{j}|.
\end{align}
According to CQA, to obtain the corresponding stationary modes $\hat{M}_{ij}$ of the physical cavity-$a$ Lindbladian, we must trace-out the ancilla mode:
\begin{align}
    \hat{M}_{ij} &=\text{Tr}_b[|\psi_{i}\rangle\langle \psi_{j}|].
\end{align}
We now compute the Wigner transform of the above stationary modes. First, we take advantage of the fact that a dark state factorizes across the two-modes $\hat{c}_\pm$ as $|\psi_{j}\rangle = |\psi_{j,+}\rangle |0_-\rangle$:
\begin{align}
\hat{M}_{ij} &=\text{Tr}_b[ (|\psi_{i,+}\rangle \langle \psi_{j,+}|)\, (|0_-\rangle \langle 0_-|)].
\end{align}
Letting $\sigma_{+,ij}(z)$ denote the symmetrically-ordered symbol (i.e. Wigner transform) of the outer-product $|\psi_{i,+}\rangle \langle \psi_{j,+}|$, and $\sigma_-(z)$ denote the symmetrically-ordered symbol of the vacuum state $|0_-\rangle \langle 0_-|$, let 
\begin{align}
    \sigma_{ij}(z)
\end{align}
denote the symmetrically-ordered symbol (i.e. the Wigner transform) of the stationary mode $\hat{M}_{ij}$. We can then rewrite the expression for the partial trace completely in terms of symmetrically-ordered symbols: in this case, the partial trace becomes an integral, and the symmetrized-antisymmetrized nature of the input states means that the integral convolves the operator symbols. The symbol $\sigma_-(z)$ then acts as a Gaussian filter for the symbol $\sigma_{+,ij}(z)$:\\
\begin{align}
\sigma_{ij}(z)&=  \int_{\mathbb{C}} d^2z' ~\sigma_-\bigg(\frac{z-z'}{\sqrt{2}}\bigg)\sigma_{+,ij}\bigg(\frac{z+z'}{\sqrt{2}}\bigg)\\
&=\int_{\mathbb{C}} d^2z_+ ~\fr{2e^{-2|z_+-\sqrt{2}z|^2}}{\pi}\sigma_{+,ij}(z_+),
\end{align}
where we have defined symmetrized and anti-symmetrized phase-space variables $z_\pm \equiv (z\pm z')/\sqrt{2}$. The above filtering operation is the same exact operation which ''reorders" a normally-ordered symbol into a symmetrically-ordered symbol, up to a rescaling of the phase space $z\mapsto \sqrt{2}z$. Indeed, we can rewrite it in terms of the heat kernel:
\begin{align}
\sigma_{ij}(z)&=  2\int_{\mathbb{C}} d^2u \,\mathcal K(t,\sqrt{2}z, u) \sigma_{+,ij}(u)\bigg|_{t=1/8}\\
&= 2\sigma^N_{+,ij}(\sqrt{2}z),\label{eq:the_key}
\end{align}
where $\sigma^N_{+,ij}$ is the normally-ordered symbol of the mode $|\psi_{i,+}\rangle \langle \psi_{j,+}|$. This form highly constrains the Wigner function of the steady state of any single-mode system with single-photon loss that is solvable via CQA.

In particular, now we can compute the symbol exactly in terms of the Segal-Bargmann representation. It is easy to show that the normally-ordered symbol of an operator has the simple form:
\begin{align}
    \sigma_A^N(z)  = \langle z|\hat{A}|z\rangle
\end{align}
where $|z\rangle$ denotes a coherent state with amplitude $z$. By expanding $\hat{A}$ in terms of outer-products as
\begin{align}
    \hat{A} = \sum_{ij}\alpha_{ij}|\psi_i\rangle \langle \psi_j|,
\end{align}
and utilizing the property $\psi_{\rm SB}(z) = \langle z^*|\psi \rangle e^{-|z|^2/2}$, Bargmann in \cite{bargmann_hilbert_1967} was able to show that this implies
\begin{align}
    \sigma_A^N(z)=\sum_{ij}\alpha_{ij}\psi_{i,\rm SB}(z^*)\psi_{j,\rm SB}(z^*)e^{-|z|^2}.
\end{align}
By substituting the exact expression for the normally-ordered symbol into Eq. (\ref{eq:the_key}), we can finally state the main result utilized in the main text:
\begin{align}
    \sigma_{ij}(z) = 2 \psi_{i,\rm SB}(\sqrt{2}z^*)\psi_{j,\rm SB}(\sqrt{2}z^*)e^{-2|z|^2}.
\end{align}
By taking linear combinations of the above stationary modes, the Wigner function of any stationary {\it density matrix} of the physical cavity $a$ has the closed-form
\begin{align}
    W_{a, ss}(z)=\fr{2}{\pi}\sum_{ij}\alpha_{ij}\psi_{i,\rm SB}(\sqrt{2}z^*)\psi_{j,\rm SB}^*(\sqrt{2}z^*)e^{-2|z|^2},
\end{align}
where $\alpha_{ij}= \alpha_{ji}^*$ is a positive semi-definite matrix with unit trace. 

We can also write the CQA ansatz in a manifestly positive-form. Letting $\{p_j\}$ denote the eigenvalues of the positive semi-definite matrix $\{\alpha_{ij}\}$, and letting $\{\phi_j(z)\}$ denote the Segal-Bargmann representations of the corresponding eigenvectors, the Wigner function can be equivalently written as
\begin{align}
    W_{a, ss}(z)=\fr{2}{\pi}\sum_{j} p_j |\phi_j(\sqrt{2}z^*)|^2e^{-2|z|^2}\geq 0,
\end{align}
where normalization forces $\sum_j p_j=1$.

As a simple example, when the dark-state subspace is one-dimensional, the Wigner function is the squared-modulus of the SB representation of the unique, normalized dark state in that subspace:
\begin{align}
    W_{a, ss}(z)=\fr{2}{\pi}|\phi(\sqrt{2}z^*)|^2e^{-2|z|^2}.
    \label{eq:dkstate_wignerfunction}
\end{align}

In summary, we have derived an exact, closed-form expression for the steady-state Wigner function of a cavity that is solvable via CQA. What is most striking from this analysis is the absence of any consideration of the Hamiltonian of the cavity: the CQA method, if it works, will predict that the Wigner function will be positive-definite, simply as a consequence of the presence of single-photon loss in the system.

\subsubsection{Other phase-space representations of the steady state}
One can also obtain the steady-state $P$ function using the results of the above analysis. Accordingly, we define the {\it generalized Weierstrass transform} $\mathcal W_t\equiv e^{-t\Delta}$ as the following integral transform:
\begin{align}
    (\mathcal W_tf)(z)\equiv \int_{\mathbb{C}} d^2u \mathcal K(t,z,u) f(u).
\end{align}
Let $(S_\alpha f)(z)\equiv f(\alpha z)$ also denote the linear operator which {\it rescales} the function argument. We then have the following commutation relation:
\begin{align}
    (S_{\sqrt{s}}\mathcal W_tf)(z)&=\int_{\mathbb{C}} d^2u \frac{e^{-\frac{|u-\sqrt{s}z|}{4t}}}{4\pi t} f(u)\nonumber\\
    &=\int_{\mathbb{C}}d^2v \frac{e^{-\frac{|v-z|^2}{4t/s}}}{4\pi t/s}f(\sqrt{s}z) = (\mathcal W_{t/s}S_{\sqrt{s}}f)(z)\label{eq:C24}
\end{align}
From Eq. \eqref{eq:C24}, it follows immediately that $\mathcal W_{t}S_{\sqrt{s}} = S_{\sqrt{s}}\mathcal W_{s\cdot t}$. With this identity in mind, we can solve for the steady-state $P$-function. Letting $Q_+,P_+$ denote the $P$- and $Q$-functions of the single-mode pure state $|\psi_+\rangle$, we have
\begin{align}
    P_{a,ss}(z) & = (\mathcal W_{-1/8}  S_{\sqrt{2}}\, Q_+)(z) =(S_{\sqrt{2}}\mathcal W_{-1/4} Q_+)(z)\nonumber\\
    &= 2 P_+(\sqrt{2}z).
\end{align}
Therefore, the $P$-function of the Kerr resonator is simply the $P$-function of the single-mode pure state $|\psi_+\rangle$, up to a rescaling of phase space $z\mapsto \sqrt{2}z$.

\section{``Gauge-invariance'' of the dark state}
\label{app:AlternateTransform}
Eq. (\ref{eq:SBTransformed2}) in the main text makes it manifest that there exist two distinct (non-unitary) gauge choices in which the troublesome two-photon term vanishes in the dark state equation Eq. (\ref{eq:NonHermKernel}). In the main text, we solved the dark state equations in the plus-gauge. This leads to the question of what would happen if we solved the dark state conditions in the minus-gauge. If we had solved for the dark state in the minus gauge, we would have ended up with the solution
\begin{align}
\widetilde{\psi}_{+,\text{SB}}(z)&\equiv e^{-\epsilon_- z}\,_1F_1\bigg(-\fr{\lambda_1+\epsilon_-D}{\epsilon_--\epsilon_+};-D;(\epsilon_--\epsilon_+)z\bigg).
\end{align}
However, the results are gauge-invariant, as we can write
\begin{align}
&\widetilde{\psi}_{+,\text{SB}}(z)= e^{-\epsilon_- z+(\epsilon_--\epsilon_+)z}\nonumber\\
&~~~~\,_1F_1\bigg(-D-\fr{\lambda_1+\epsilon_-D}{\epsilon_+-\epsilon_-};-D;(\epsilon_+-\epsilon_-)z\bigg)\label{eq:kummer_transform}\\
&= e^{-\epsilon_+z}\nonumber\\
&~~~~\,_1F_1\bigg(-\fr{(\epsilon_+-\epsilon_-)D}{\epsilon_+-\epsilon_-}-\fr{\lambda_1+\epsilon_-D}{\epsilon_+-\epsilon_-};-D;(\epsilon_+-\epsilon_-)z\bigg)\nonumber\\
&= e^{-\epsilon_+z}\,_1F_1\bigg(-\fr{\lambda_1+\epsilon_+D}{\epsilon_+-\epsilon_-};-D;(\epsilon_+-\epsilon_-)z\bigg)=\psi_{+,\rm SB},
\end{align}
where $\psi_{+,\rm SB}$ is the dark state Eq. (\ref{eq:Kumm2}) in the main text, and in the first line (Eq. (\ref{eq:kummer_transform})) we utilized {\it Kummer's transformation} (see e.g. Ref.~\cite{brychkov_handbook_nodate}), which is a fundamental symmetry of the confluent hypergeometric differential equation:
\begin{align}
    \,_1F_1(r_1;r_2;z)&=e^{z}\,_1F_1(r_2-r_1;r_2;-z).\\
    \nonumber
\end{align}

\section{Beyond nonlinear single-photon driving: breakdown of CQA}

The most general Kerr Hamiltonian, that is, containing all possible terms of lower order than the Kerr nonlinearity, is

\begin{align}
     \hat{H}_a &=
        \fr{K}{2}\hat{a}^\dag \hat{a}^\dag \hat{a}\hat{a} -\Delta \hat{a}^\dag \hat{a}
\nonumber        \\
        &+ \bigg[ 
            \bigg(\Lambda_1\hat{a}^\dag + \fr{\Lambda_2}{2}\hat{a}^\dag \hat{a}^\dag+\Lambda_3 \hat{a}^\dag \hat{a}^\dag \hat{a}+\Lambda_4\, \hat{a}^\dagger\hat{a}^\dagger\hat{a}^\dagger \bigg) + h.c. \bigg]  . \label{eq:ciuti_ham}
\end{align}
This begs the question of why everything of degree three or lower is exactly solvable by CQA, except the $(\hat{a}^\dag)^3$ term. The explanation for this is rather simple: when acting the cascaded Hamiltonian on the dark state ansatz, one gets the equation
\begin{align}
    (\hat{c}_-^\dag\mathcal{\hat{H}}_++\frac{\Lambda_4}{\sqrt{8}}(\hat{c}_-^\dag)^3)|\psi_+\rangle=0\label{eq:dk3ph}
\end{align}
where
\begin{align}
    \mathcal{\hat{H}}_+ &\to \mathcal{\hat{H}}_+ +\frac{3\Lambda_4}{\sqrt{8}}(\hat{c}_+^\dag)^2.
\end{align}
The shift in $\hat{\mathcal{H}}_+$ is innocuous. However, the term cubic in $\hat{c}_-^\dag$ is lethal: both terms in Eq. \eqref{eq:dk3ph} have respectively one and three photons in the minus mode, and are thus generically orthogonal. Thus the only solution to the dark state condition Eq. \eqref{eq:dk3ph} is $|\psi_+\rangle = 0$.


\section{Stationary density matrix and moments of a driven Kerr cavity}\label{app:formal}

We show here how to compute exact analytic expressions from our steady-state solution for density matrix elements in the Fock basis, as well as normal-ordered cavity moments. Although the expressions obtained here are considerably more complex/physically opaque, this will allow us to make contact with older results obtained via $P$-function methods \cite{bartolo_exact_2016, elliott_applications_2016}. Expanding the purification of the density matrix $|\psi\rangle=|\psi_+\rangle\otimes |0_-\rangle$, and writing the symmetric component in the SB representation yields
\begin{align}
\psi_{+,\rm SB}(z)\equiv \sum_{l=0}^\infty \psi_l\fr{z^l}{l!},
\end{align}
which implicitly defines coefficients $\psi_l\equiv \psi_{+,\rm SB}^{(l)}(0)$ which are the derivatives of the Bargmann state evaluated at the origin $z=0$ in phase space. In the special cases $\lambda_1\equiv \lambda_3 \equiv  0$, reproducing results in \cite{bartolo_exact_2016}), or the more generic regime $\lambda_2 \equiv 0$, which represents new results, we can actually evaluate the sums, resulting in compact, closed-form expressions.

\subsubsection{Steady-state density matrix}
In terms of these Taylor coefficients, the steady state density matrix can be computed in the Fock basis:
\begin{align}
&\langle m|\hat{\rho}_{a,ss}|n\rangle \\
&= \sum_{l=0}^\infty \langle m,l|\bigg(\sum_{j,k=0}^\infty \fr{\psi_j\psi_k^*}{j!k!}(\hat{c}_+^\dag)^j|0\rangle \langle 0|\hat{c}_+^k\bigg)|n,l\rangle\nonumber\\
&=\fr{1}{(2^{m+n}n!m!)^{1/2}}\nonumber\\
&\sum_{l,j,k=0}^\infty \fr{\psi_j\psi_k^*}{j!k!}\langle 0|\fr{(\hat{c}_+-\hat{c}_-)^{l}}{\sqrt{2^ll!}}(\hat{c}_++\hat{c}_-)^m(\hat{c}_+^\dag)^j|0\rangle\nonumber\\
&~~~~~~~~~~~~~~~~~~~~~~~~~\cdot\langle 0|\hat{c}_+^k(\hat{c}_+^\dag+\hat{c}_-^\dag)^{n}\fr{(\hat{c}_+^\dag-\hat{c}_-^\dag)^l}{\sqrt{2^ll!}}|0\rangle \nonumber\\
&=\fr{1}{(2^{m+n}n!m!)^{1/2}}\nonumber\\
&\sum_{j,k,l=0}^\infty \fr{\psi_j\psi_k^*}{j!k!}\fr{1}{2^ll!}\langle 0|\hat{c}_+^{m+l}(\hat{c}_+^\dag)^j|0\rangle\langle 0|\hat{c}_+^k(\hat{c}_+^\dag)^{n+l}|0\rangle .
\end{align}
Using identities of the form $\langle 0|\hat{c}_+^{m+l}(\hat{c}_+^\dag)^j|0\rangle=\delta_{m+l,j}j!$, etc., we get the remarkably simple result:\\
\begin{align}
\langle m|\hat{\rho}_{a,ss}|n\rangle=\fr{1}{\sqrt{2^{m+n}n!m!}}\sum_{l=0}^\infty \fr{\psi_{m+l}\psi_{n+l}^{*}}{2^ll!}.
\end{align}
This expression matches similar expressions obtained using complex-P solutions, as we will see later in this section.

\subsubsection{Cavity moments}
We can also express the normally-ordered moments of a driven Kerr cavity exactly in terms of the scaled Fock-state amplitudes $\psi_l$. The calculation is slightly more straightforward:
\begin{align}
\text{Tr}[ \hat{\rho}_{a,ss}(a^\dag)^n a^m]&=\langle \psi|(a^\dag)^na^m|\psi\rangle\nonumber\\
&=\fr{1}{\sqrt{2^{m+n}}}\langle \psi|(\hat{c}_+^\dag+\hat{c}_-^\dag)^n (\hat{c}_++\hat{c}_-)^m|\psi\rangle,
\end{align}
where $|\psi\rangle$, as before, is the purification of the density matrix obtained from the absorber method. Expanding the dark state yields
\begin{align}
&\text{Tr}[ \hat{\rho}_{a,ss}(a^\dag)^n a^m]\nonumber\\
&=\fr{1}{\sqrt{2^{m+n}}}\sum_{j,k=0}^\infty\fr{\psi_j^*\psi_k}{j!k!}\langle 0|\hat{c}_+^j(\hat{c}_+^\dag)^n \hat{c}_+^m(\hat{c}_+^\dag)^k|0\rangle\nonumber\\
&=\fr{1}{\sqrt{2^{m+n}}}\sum_{j,k=0}^\infty \fr{\psi_j^*\psi_k}{\sqrt{j!k!}}\langle j_+|(\hat{c}_+^\dag)^n \hat{c}_+^m|k_+\rangle. 
\end{align}
Defining a new variable $l$ such that $j\equiv n+l$, we find that $k=m+l$, and that furthermore $l\geq 0$. So our sum simplifies to 
\begin{align}
&\text{Tr}[ \hat{\rho}_{a,ss}(a^\dag)^n a^m]\nonumber\\
&\fr{1}{\sqrt{2^{m+n}}}\sum_{l=0}^\infty \fr{\psi_{n+l}^*\psi_{m+l}}{\sqrt{(m+l)!(n+l)!}}\fr{\sqrt{(m+l)!}}{\sqrt{l!}} \fr{\sqrt{(n+l)!}}{\sqrt{l!}}.
\end{align}
We thus obtain the simple result
\begin{align}
    \text{Tr}[ \hat{\rho}_{a,ss}(a^\dag)^n a^m]&=\fr{1}{\sqrt{2^{m+n}}}\sum_{l=0}^\infty \fr{\psi_{m+l}\psi_{n+l}^*}{l!}.
\end{align}
This is the formula used to produce exact-solution plots of average photon number in Fig.~\ref{fig:blockades}; a similar-looking expression was derived independently in \cite{bartolo_exact_2016}, using $P$-function methods.

\subsubsection{Normalization}
Throughout this section, we have assumed that the normalization of $|\psi_+\rangle$ is known. Supposing that this is {\it not} the case, and $|\psi_+\rangle$ is written instead in the form
\begin{align}
\psi_{+,\rm SB}(z)=\fr{1}{\sqrt{N}}\sum_{l=0}^\infty \widetilde{\psi_l} \fr{z^l}{l!},
\end{align}
we can write an exact expression for $N$:
\begin{align}
N=\sum_{l=0}^\infty \fr{\widetilde{\psi_l}}{\sqrt{l!}}
 \fr{\widetilde{\psi_l}^*}{\sqrt{l!}}=\sum_{l=0}^\infty \fr{|\widetilde{\psi_l}|^2}{l!}.\label{eq:normalization}
\end{align}

\subsubsection{Expression for $\psi_l$ in general regime}
The scaled Fock-state amplitudes $\psi_l$ can be computed in closed-form in terms of the Gauss hypergeometric function. We can then utilize this closed form to show that our exact expressions derived here agree with earlier solutions \cite{bartolo_exact_2016, elliott_applications_2016} in the limit of $\Lambda_3\to 0$:
\begin{align}
\psi_l&\equiv \partial^l\psi_{+,\rm SB}(0).
\end{align}
The above quantity is particularly difficult to evaluate in the general case, so we evaluate instead
\begin{align}
\xi_l&\equiv \partial^l\xi_{+,\rm SB}(0).\label{eq:displ_psil}
\end{align}
where $|\xi\rangle$ is the displaced dark state in the main text. Eq. (\ref{eq:displ_psil}) then represents the Fock-state amplitudes of the purification of the {\it displaced} steady-state $\hat{\rho}'\equiv \hat{D}_\alpha \hat{\rho}_{a,ss} \hat{D}_\alpha^\dag$, where $\alpha$ is defined in the main text and vanishes when $\Lambda_3\to 0$.\\

Expanding $\xi_{+,\rm SB}(z)\equiv \Theta(z) \phi(z)$, where $\Theta(z)\equiv \exp(-\theta(z))$ is the non-unitary gauge transformation in the main text. Expanding via the Leibniz rule, we get
\begin{align}
\xi_l&=\sum_{n=0}^l  \binom{l}{n}\partial^{l-n}\Theta(0)\partial^n\phi(0).
\end{align}
Plugging in $\Theta(z)\equiv e^{-\epsilon_+ z}$ and $\phi(z) = \,_1F_1(-r_1;-r_2;(\epsilon_+-\epsilon_-) z)$, we get
\begin{align}
\partial^k\Theta(0)&=(-\epsilon_+)^k\\
\partial^k\phi(0)&=\fr{(-r_1)_k}{(-r_2)_k}(\epsilon_+-\epsilon_-)^k
\end{align}
So, in total, we get
\begin{align}
\xi_l&=\sum_{n=0}^l \binom{l}{n}\fr{(-r_1)_n}{(-r_2)_n}(-\epsilon_+)^{l-n}(\epsilon_+-\epsilon_-)^n\nonumber\\
&=(-\epsilon_+)^l\sum_{n=0}^l(-1)^n \binom{l}{n}\fr{(-r_1)_n}{(-r_2)_n}\bigg(1-\fr{\epsilon_-}{\epsilon_+}\bigg)^n\nonumber\\
&=(-\epsilon_+)^l\sum_{n=0}^l\fr{(-l)_n(-r_1)_n}{(-r_2)_n}\fr{(1-\fr{\epsilon_-}{\epsilon_+})^n}{n!}.\nonumber\\
&=(-\epsilon_+)^l\sum_{n=0}^\infty\fr{(-l)_n(-r_1)_n}{(-r_2)_n}\fr{(1-\fr{\epsilon_-}{\epsilon_+})^n}{n!}.
\end{align}
Therefore, we have a closed-form expression for the scaled Fock-state amplitudes of the displaced dark state:
\begin{align}
\xi_l&=(-\epsilon_+)^l\,_2F_1(-l,-r_1;-r_2;1-\tfrac{\epsilon_-}{\epsilon_+}).
\end{align}
In the limit $\Lambda_3\to 0$, $\epsilon_+\to -\epsilon_-$, and so, as in the main text, defining $\epsilon \equiv \epsilon_+$,  we get
\begin{align}
\psi_l&\underset{\Lambda_3\to 0}{\sim}(-\epsilon)^l\,_2F_1(-l,-r_1;-r_2;2).\label{eq:recover}
\end{align}
where we are implicitly utilizing the fact that $\xi_l \to \psi_l$ in this limit, as the displacement parameter $\alpha$ vanishes in the limit $\Lambda_3\to 0$. From Eq. (\ref{eq:recover}), it is straightforward to recover the previous solutions \cite{bartolo_exact_2016, elliott_applications_2016} of the Kerr resonator in the limit $\Lambda_3\equiv 0$.

\section{Exact results when the non-unitary gauge transformation is trivial}
\label{app:l2}

The series expressions derived in Appendix \ref{app:formal} have simple closed forms when we have $\lambda_2 \equiv 0$, which, for $\Lambda_2\neq 0$, represents previously unexplored physics. We emphasize the generic nature of this regime, in that there are 8 real parameters to play with: $\Lambda_1,\Lambda_3$, and $K,\Delta,\kappa_1,\kappa_2$. In this limit, the displaced SB wavefunction Eq. (\ref{eq:displacement}) is purely hypergeometric:
\begin{align}
\xi_{+,\rm SB}(z)&=\fr{1}{N^{1/2}}\,_1F_1(-r_1;-r_2;-\lambda_3z),\label{eq:dark_state_no_gauge}
\end{align}
where $r_1\equiv -\lambda_1/\lambda_3$, and $r_2\equiv D$. In this case, the coefficients $\xi_l$ of the displaced steady state simplify to ratios of Pochhammer symbols:
\begin{align}
\xi_l=\fr{1}{N^{1/2}}\fr{(-r_1)_l}{(-r_2)_l}(-\lambda_3)^l,
\end{align}
where the Pochhammer symbol is defined as $(z)_l\equiv \Gamma(z+l)/\Gamma(z)$. Therefore, the normalization is computable in closed form: 
\begin{align}
N&=\sum_{l=0}^\infty \fr{(-r_1)_l(-r_1^*)_l}{(-r_2)_l(-r_2^*)_l}\fr{|\lambda_3|^{2l}}{l!}\nonumber\\
&~~~~~~~~~~~~~=\,_2F_2(-r_1,-r_1^*;-r_2,-r_2^*;|\lambda_3|^2).
\end{align}
Here,  $\,_pF_q(a_1\cdots a_p; b_1,\cdots b_q;z)$ denotes the generalized hypergeometric function (see, e.g. \cite{brychkov_handbook_nodate}). The normalization of the steady-state Wigner function is thus exactly computable:
\begin{align}
W_{a,ss}(z-\alpha)&=\fr{2|_1F_1(-r_1;-r_2;-\sqrt{2}\lambda_3z^*)|^2e^{-2|z|^2}}{\pi\,_2F_2(-r_1,-r_1^*;-r_2,-r_2^*;|\lambda_3|^2)^{1/2}},
\end{align}
where here $\alpha \equiv \alpha_+/\sqrt{2}$ is the appropriately normalized displacement factor given in Eq. (\ref{eq:displacement}) in the main text. We now move on to compute the matrix elements of the density matrix in the displaced frame (here, $ \hat{D}_\alpha\equiv e^{-\alpha \hat{a}^\dag - h.c.}$ is the standard displacement operator of the physical cavity).\\
\begin{align}
&\langle m|\hat{D}_\alpha\hat{\rho}_{a,ss}\hat{D}_\alpha^\dag|n\rangle=\fr{(-\lambda_3)^m(-\lambda_3^*)^n}{N\sqrt{2^{m+n}n!m!}}\nonumber\\
&~~~~~~~~~~~~~~~~~~~~~~~~~\cdot \sum_{l=0}^\infty \fr{(-r_1)_{m+l}(-r_1^*)_{n+l}}{(-r_2)_{m+l}(-r_2^*)_{n+l}}\fr{(|\lambda_3|^{2}/2)^l}{l!}.\nonumber
\end{align}
Utilizing the identity $(z)_{m+l}=(z)_m(z+m)_l$, the sum closes, and we get
\begin{align}
&\langle m|\hat{D}_\alpha\hat{\rho}_{a,ss}\hat{D}_\alpha^\dag|n\rangle=\fr{\xi_m\xi_n^*}{\sqrt{2^{m+n}n!m!}}\nonumber\\
&\cdot \,_2F_2(n-r_1,m-r_1^*;m-r_2;n-r_2^*;|\lambda_3|^2/2)\label{eq:l2_photon_number_distribution}
\end{align}
As for the normally-ordered cavity moments in the displaced frame, in a similar fashion, we get an analogous closed-form in terms of a generalized hypergeometric function:
\begin{align}
&\text{Tr}[\hat{D}_\alpha \hat{\rho}_{a,ss}\hat{D}_\alpha^\dag(\hat{a}^\dag)^n\hat{a}^m] = \fr{\xi_m\xi_n^*}{\sqrt{2^{m+n}}}\nonumber\\
& \cdot\,_2F_2(m-r_1,n-r_1^*;m-r_2;n-r_2^*;|\lambda_3|^2)
\end{align}

\section{Exact results in the parity-conserving regime}
\label{app:TracingOut}
We now will complete the process started in Section \ref{sec:bistability}, namely that of tracing-out the ancilla resonator for each of the dark steady states obtained by the CQA method. We begin with the formula in Appendix \ref{app:formal} on unique steady states:
\begin{align}
\langle m|\hat{\rho}_{a,ss}|n\rangle&=\fr{1}{\sqrt{2^{m+n}m!n!}}\sum_{l=0}^\infty \fr{\psi_{m+l}\psi_{n+l}^{*}}{2^ll!}
\end{align}
note that, as a direct consequence of $\psi_{2l-1}\equiv 0$, we have\\
\begin{align}
\langle 2j+1|\hat{\rho}_{a,ss}|2k\rangle=\langle 2j|\hat{\rho}_{a,ss}|2k+1\rangle=0,
\end{align}
as each term in the sum over $l$ would identically vanish in these cases. In summary,
\begin{align}
    \hat{\Pi}_e\hat{\rho}_{a,ss}\hat{\Pi}_o=\hat{\Pi}_o\hat{\rho}_{a,ss}\hat{\Pi}_e=0,
\end{align}
where $\hat{\Pi}_{e/o}$ are the projections onto the subspaces of the resonator Hilbert space spanned by even/odd photon number states.

Therefore, 
by taking matrix elements on both sides of Eq. (\ref{eq:bistable_manifold}) in the main text, one obtains
\begin{align}
\langle m|\hat{\rho}_e|n\rangle&=\fr{2N}{N+1}\langle m|\hat{\rho}_{a,ss}|n\rangle,~~~m,n~~\text{even}\label{eq:even_ss}\\
\langle m|\hat{\rho}_o|n\rangle&=\fr{2N}{N-1}\langle m|\hat{\rho}_{a,ss}|n\rangle ,~~~m,n~~\text{odd}\label{eq:odd_ss}.
\end{align}
Therefore, to compute the steady states $\hat{\rho}_{e/o}$, it suffices to compute matrix elements of $\hat{\rho}_{a,ss}$. We note that this was done in \cite{bartolo_exact_2016} (as this represents the unique steady-state regime $\kappa_1\neq 0$), and so we're technically done, as we could simply cite the result here.

 \begin{figure}
      \centering
      \includegraphics[width=0.9\columnwidth]{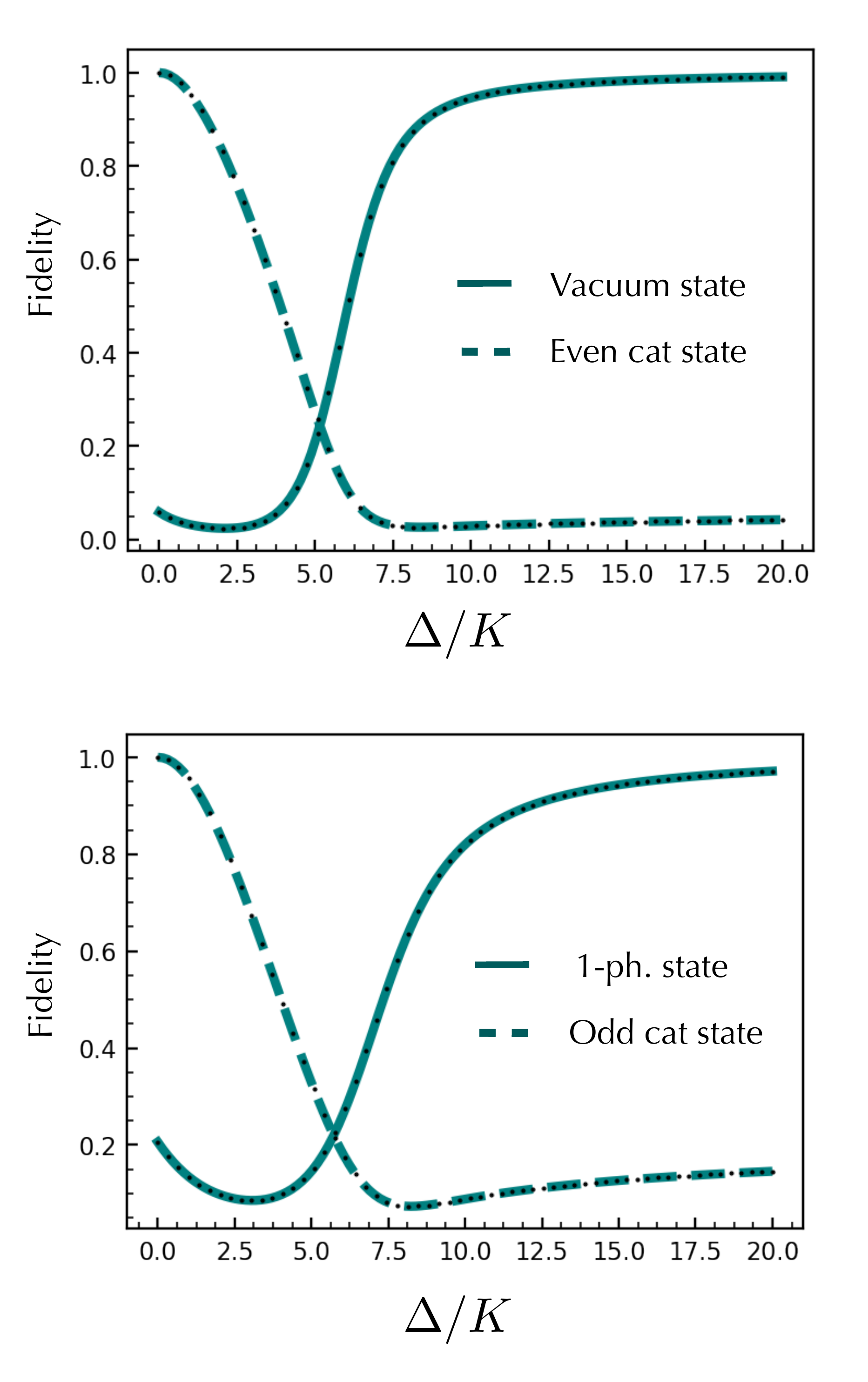}
      \caption{{\bf Limiting behavior of bistable states}. (a). We plot the fidelity of $\hat{\rho}_e$ (c.f. Eq. (\ref{eq:even_ss})) with an even cat state with amplitude $\alpha =i\sqrt{\lambda_2/2}$ (solid line) and the vacuum state (dashed line). Corresponding results from exact diagonalization are also given (black dots) (b) We plot the fidelity of $\hat{\rho}_o$ (c.f. Eq. (\ref{eq:odd_ss})) with an amplitude-$\alpha$ odd cat state  (solid line) and a 1-photon Fock state (dashed line). Corresponding results from exact diagonalization are also given (black dots). {\it Parameter choices}: In both plots, $\Lambda_2=5K$, $\kappa_2=K$, and $\Lambda_1,\kappa_1\equiv 0$.}
      \label{fig:fidelity}
  \end{figure}

For completeness, however, we show that the calculation of the expressions on the RHS's of Eqs.  (\ref{eq:even_ss}-\ref{eq:odd_ss}) can be reproduced in a straightforward manner within the quantum absorber formalism. Assuming $m\equiv 2j$, $n\equiv 2k$ are both even, we have
\begin{align}
&\sum_{l=0}^\infty\fr{\psi_{m+l}\psi_{n+l}^{*}}{2^ll!}=\sum_{l=0}^\infty\fr{\psi_{2(j+l)}\psi_{2(k+l)}^{*}}{2^{2l}(2l)!}\nonumber\\
&=\psi_{2j}\psi_{2k}^*\sum_{l=0}^\infty\fr{(j+\fr{1}{2})_l(k+\fr{1}{2})_l}{(j-r_2)_l(k-r_2^*)_l}\fr{|\lambda_2|^{2l}}{2^{2l}(2l)!}\nonumber\\
&=\psi_{2j}\psi_{2k}^*\sum_{l=0}^\infty\fr{(j+\fr{1}{2})_l(k+\fr{1}{2})_l}{(j-r_2)_l(k-r_2^*)_l(\fr{1}{2})_l}\fr{|\lambda_2/4|^{2l}}{l!}
\end{align}
Therefore, in total we have
\begin{align}
&\sum_{l=0}^\infty\fr{\psi_{m+l}\psi_{n+l}^{*}}{2^ll!}\\
&~~~~=\psi_{2j}\psi_{2k}^*\cdot \pFq[20]{2}{3}{j+\fr{1}{2},~k+\fr{1}{2}~~~~~~~}{j-r_2,k-r_2^*, \fr{1}{2}}{|\lambda_2/4|^2}.
\end{align}
Assuming $m\equiv 2j+1$, $n\equiv 2k+1$ are both odd, we have
\begin{align}
&\sum_{l=0}^\infty\fr{\psi_{m+l}\psi_{n+l}^{*}}{2^ll!}=\sum_{l=0}^\infty\fr{\psi_{2(j+l+1)}\psi_{2(k+l+1)}^{*}}{2^{2l+1}(2l+1)!}\nonumber\\
&=4\sum_{l=1}^\infty \fr{l\psi_{2(j+l)}\psi_{2(k+l)}^*}{2^{2l}(2l)!}\nonumber\\
&=4\psi_{2j}\psi_{2k}^*\sum_{l=1}^\infty \fr{l(\fr{1}{2}+j)_l(\fr{1}{2}+k)_l}{(j-r_2)_l(k-r_2^*)_l(\fr{1}{2})_l}\fr{|\lambda_2/4|^{2l}}{l!}\nonumber\\
&=4\psi_{2j}\psi_{2k}^*\fr{(j+\fr{1}{2})(k+\fr{1}{2})|\lambda_2/4|^2}{(j-r_2)(k-r_2^*)(\fr{1}{2})}\nonumber \\
&~~~~\cdot\sum_{l=1}^\infty \fr{(\fr{3}{2}+j)_l(\fr{3}{2}+k)_l}{(j+1-r_2)_l(k+1-r_2^*)_l(\fr{3}{2})_l}\fr{|\lambda_2/4|^{2l}}{l!}
\end{align}
Therefore, in total we have
\begin{align}
&\sum_{l=0}^\infty\fr{\psi_{m+l}\psi_{n+l}^{*}}{2^ll!}=\fr{\psi_{2j}\psi_{2k}^*}{2}\fr{(j+\fr{1}{2})(k+\fr{1}{2})|\lambda_2|^2}{(j-r_2)(k-r_2^*)}\nonumber \\
&~~~~~~~~~~~~~~\cdot \pFq[20]{2}{3}{j+\fr{3}{2},~k+\fr{3}{2}~~~~~~~}{j+1-r_2,k+1-r_2^*, \fr{3}{2}}{|\lambda_2/4|^2}.
\end{align}
In summary, we have the following closed-form for $\hat{\rho}_{a,ss}$ in the Fock basis:
\begin{align}
&\langle m|\hat{\rho}_{a,ss}|n\rangle\underset{m,n\in 2\mathbb{Z}}{=}\fr{\psi_m\psi_n^*}{\sqrt{2^{m+n}m!n!}}\nonumber\\
&~~~~~~~~~~~~~~~~~~~~~~~~\pFq[20]{2}{3}{\fr{m+1}{2},~\fr{n+1}{2}}{\fr{m-D+1}{2},\fr{n-D^*+1}{2}, \fr{1}{2}}{\bigg|\frac{\lambda_2}{4}\bigg|^2},\label{eq:rho_even_exact}\\
&\langle m|\hat{\rho}_{a,ss}|n\rangle\underset{m,n\in 1+2\mathbb{Z}}{=}\fr{\psi_{m-1}\psi_{n-1}^*}{\sqrt{2^{m+n}m!n!}}\nonumber \\
&\fr{|\lambda_2|^2mn}{2(m-D)(n-D^*)}\cdot \pFq[20]{2}{3}{\fr{m+2}{2},~\fr{n+2}{2}}{\fr{m-D+2}{2},\fr{n-D^*+2}{2}, \fr{3}{2}}{\bigg|\frac{\lambda_2}{4}\bigg|^2}.\label{eq:rho_odd_exact}
\end{align}
Substituting Eq.'s (\ref{eq:rho_even_exact}-\ref{eq:rho_odd_exact})
into Eq.'s (\ref{eq:even_ss}-\ref{eq:odd_ss}) , we get that the bistable manifold of the Kerr cavity in this regime is spanned by the following density matrices:
\begin{align}
&\langle m|\hat{\rho}_{e}|n\rangle\underset{m,n\in 2\mathbb{Z}}{=}\fr{2N}{N+1}\fr{\psi_m\psi_n^*}{\sqrt{2^{m+n}m!n!}}\nonumber\\
&~~~~~~~~~~~~~~~~~~~~~~~~\cdot \pFq[20]{2}{3}{\fr{m+1}{2},~\fr{n+1}{2}}{\fr{m-D+1}{2},\fr{n-D^*+1}{2}, \fr{1}{2}}{\bigg|\frac{\lambda_2}{4}\bigg|^2},\\
&\langle m|\hat{\rho}_{o}|n\rangle\underset{m,n\in 1+2\mathbb{Z}}{=}\fr{2N}{N-1}\fr{\psi_{m-1}\psi_{n-1}^*}{\sqrt{2^{m+n}m!n!}}\nonumber \\
&\fr{|\lambda_2|^2mn}{2(m-D)(n-D^*)} \pFq[20]{2}{3}{\fr{m+2}{2},~\fr{n+2}{2}}{\fr{m-D+2}{2},\fr{n-D^*+2}{2}, \fr{3}{2}}{\bigg|\frac{\lambda_2}{4}\bigg|^2},
\end{align}
and with all other matrix elements vanishing. Here, $N$ has the closed-form expression (also given in Eq. (\ref{eq:normalization2})):
\begin{align}
    N=\,_1F_2(1/2; 1/2-D/2,(1/2-D/2)^*;|\lambda_2/2|^2).\label{eq:normalization_app}
\end{align}
For small detuning $|D|\ll 1$, the CQA solutions above approach even/odd cat states, both of which exhibit Wigner function negativity. In the large detuning limit $|D|\gg 1$, $\hat{\rho}_{e/o}$ also approach pure states:  the vacuum state and one-photon state respectively, one of which exhibits Wigner function negativity. The exact CQA solutions are validated against master equation numerics in Figure \ref{fig:fidelity}.

For completeness, we include here the calculation of $N$ in Eq. (\ref{eq:normalization_app}) (this expression also shows up in the main text, in Eq. (\ref{eq:bistable_manifold}), where it controls the average parity of the unique steady state selected when parity-symmetry is spontaneously broken). The derivatives of the dark state, evaluated at the origin in phase space, are
\begin{align}
\psi_{2l}&=\fr{1}{N^{1/2}}\fr{(2l)!}{l!(\fr{1}{2}-\fr{D}{2})_l}(-\lambda_2/4)^l
\end{align}
whereas the odd derivatives vanish at the origin ($\psi_{2l-1}=0$). However, note the following identity $\fr{(2l)!}{l!}=2^{2l}(\fr{1}{2})_l$. With this identity, the Fock state amplitudes take the simpler form:
\begin{align}
\psi_{2l}&=\fr{1}{N^{1/2}}\fr{(-r_1)_l}{(-r_2)_l}(-\lambda_2)^l\label{eq:psi2l}
\end{align}
with $r_1\equiv -1/2$, and $r_2\equiv r_1+D/2$. Having computed the Taylor coefficients $\psi_l$, Eq. (\ref{eq:normalization}) gives us the normalization:
\begin{align}
N&=\sum_{l=0}^\infty \fr{(\fr{1}{2})_l(\fr{1}{2})_l}{(-r_2)_l(-r_2^*)_l}\fr{|\lambda_2|^{2l}}{(2l)!}\nonumber\\
&=\sum_{l=0}^\infty \fr{(\fr{1}{2})_l}{(-r_2)_l(-r_2^*)_l}\fr{|\lambda_2/2|^{2l}}{l!}\nonumber\\
&~~~~~~~~~~~~~=\,_1F_2(\tfrac{1}{2};-r_2,-r_2^*;|\lambda_2|^2).
\end{align}
This concludes the calculation of the normalization constants in the expressions Eqs. (\ref{eq:even_ss}-\ref{eq:odd_ss}).

\bibliography{kerr_bistability}

\end{document}